# Interactive introduction to self-calibrating interfaces

A self-calibrating interface can identify what you are trying to do without knowing how you are trying to do it. In this article, you will discover and interact with a self-calibrating PIN-entry interface and visualize how it works.

Jonathan Grizou
jonathan.grizou@glasgow.ac.uk

*School of computing Science, University of Glasgow, Scotland, UK*
*INSERM U1284, Université de Paris, Center for Research and Interdisciplinarity (CRI), Paris, France*

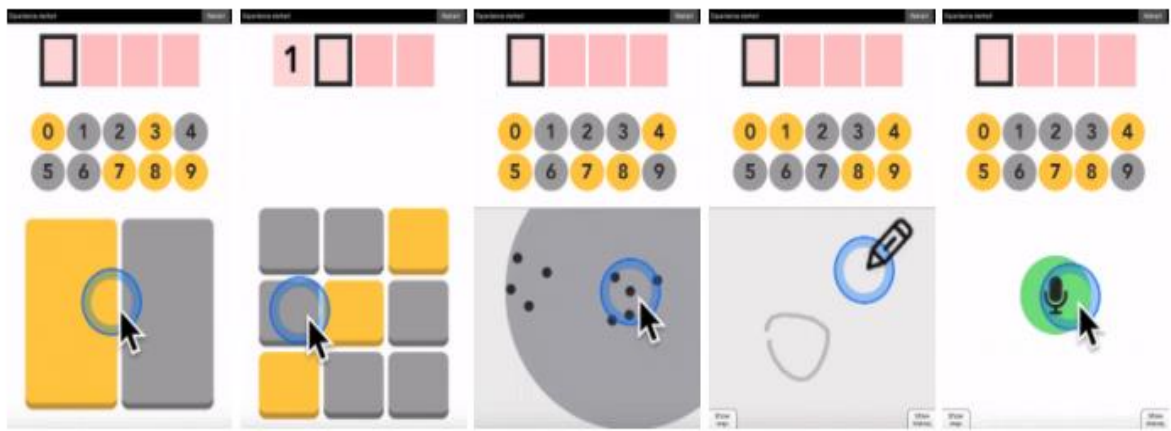

## Abstract

This interactive paper aims to provide an intuitive understanding of the self-calibrating interface paradigm. Under this paradigm, you can choose how to use an interface which can adapt to your preferences on the fly. We introduce a PIN entering task and gradually release constraints, moving from a pre-calibrated interface to a self-calibrating interface while increasing the complexity of input modalities from buttons, to points on a map, to sketches, and finally to spoken words. This is not a traditional research paper with a hypothesis and experimental results to support claims; the research supporting this work has already been done and we refer to it extensively in the later sections. Instead, our aim is to walk you through an intriguing interaction paradigm in small logical steps with supporting illustrations, interactive demonstrations, and videos to reinforce your learning. We designed this paper for the enjoyments of curious minds of any backgrounds, it is written in plain English and no prior knowledge is necessary. All demos are available online at openvault.jgrizou.com and linked individually in the paper.

# Table of Contents

# Introduction

Interfaces are all around us. A keyboard, the code-pad of an ATM, the touch screen of a vending machine, the remote control of a TV are all interfaces. They sit between you and the machine you want to control. Their function is to translate your button presses into information a machine can understand and act upon.

Letters or symbols are printed on these buttons to tell you what they do. When you press the letter printed 'A' on your keyboard, the machine assumes that you want to type the letter A and acts accordingly.

In this article we question the need to print explanatory symbols on buttons. Could we use buttons the way we want, instead of having to learn which ones to use? Could a machine understand what we want to achieve, without knowing how to interpret our actions?

We will show that this is possible under specific, but not uncommon, conditions. To demonstrate this, we designed a PIN-entering interface that you will use to type a 4-digit PIN of your choice.

We will progress through 5 different versions of this interface. The first version will use standard buttons whose color will indicate what they do. The second version will not have any pre-assigned color on each button. You will assign colors to buttons in your mind, without disclosing it to the machine, and the interface will nonetheless understand you. The subsequent versions will go further and remove buttons altogether, replacing them by points on a 2D map, hand drawn symbols and, finally, simple vocal commands in a language you will invent.

To start, you should understand how to use our PIN-entering interface. We call this interface IFTT-PIN, for "IF This Then PIN". It is a bit different from a typical code-pad. Instead of pressing directly on digits, the digits are color-coded. To type a digit, you will press colored buttons to inform the machine of the color associated with your digit.

Our interface can be broken down into three parts (Figure 1). The top part is where your 4-digit PIN will be displayed. The middle part shows all digits from 0 to 9 colored in either yellow or grey. The bottom part is for user interaction, for now it contains two buttons, one yellow and one grey.

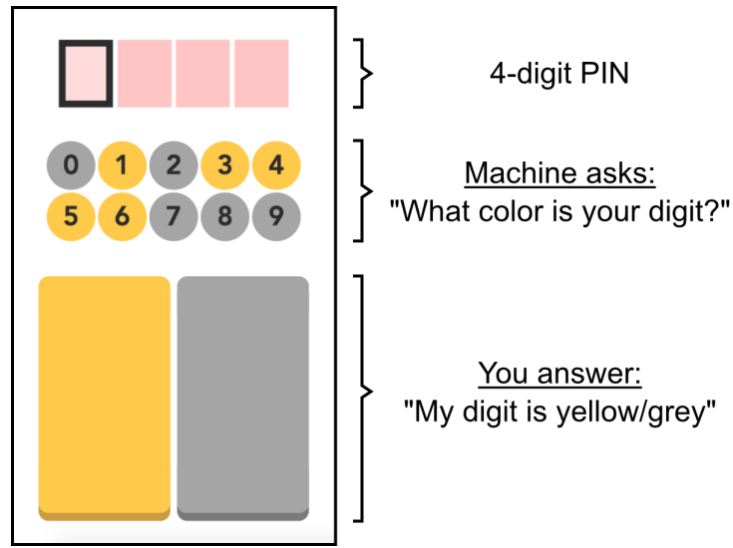

Figure 1: Breakdown of the IFTT-PIN interface

Think of the machine as asking you: "What color is your digit?". Looking at the digit you want to enter, you answer: "My digit is yellow" or "My digit is grey" by clicking on the corresponding button.

For example, if your PIN is 1234. You would start by typing the first digit of our PIN[1], which is a 1. Because the digit 1 is currently colored in yellow (Figure 1), you would click on the yellow button. The machine could then immediately discard all the grey digits and infer that the first digit of your PIN is either 1, 3, 4, 5, or 6 (Figure 2).

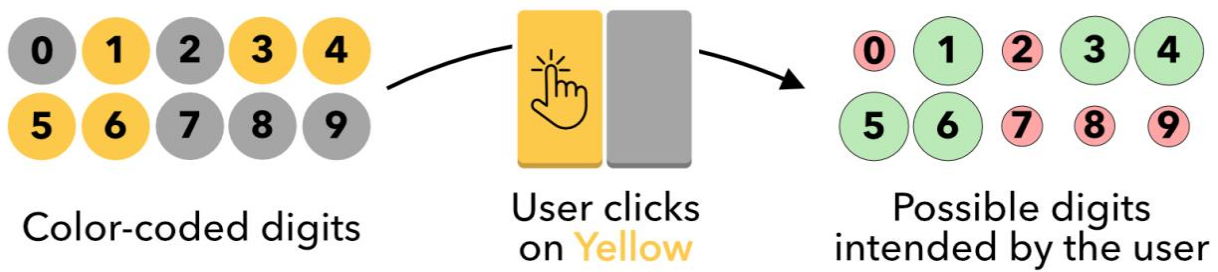

Figure 2: Clicking on the yellow reduces the possible digit to 1, 3, 4, 5 or 6

---

[1] As indicated by the dark frame around the leftmost digit in the top of Figure 1

The digit's colors will then change and by repeating the process a few times the machine will identify your first digit with certainty (Figure 3).

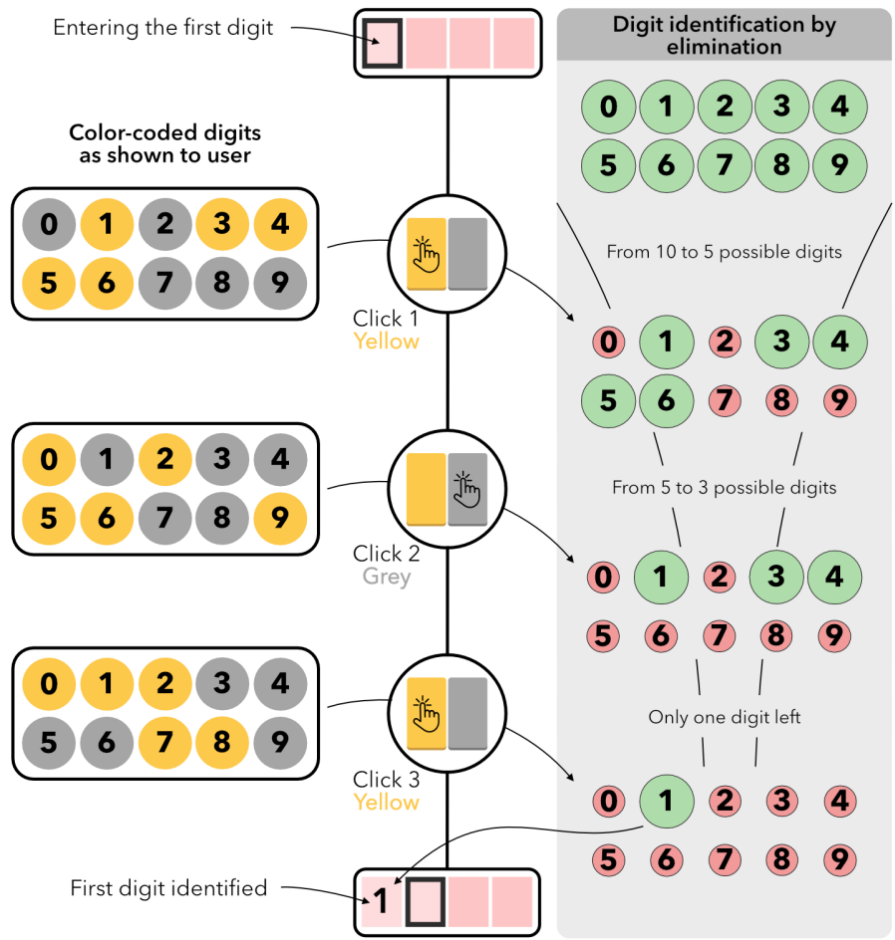

*Figure 3: Digit identification process. Left: Color-coded digits as visible to the user. Middle: Action taken by the user. Right: Digit elimination process inside the machine.*

Once the first digit is identified, we can repeat the same process for the second digit, then the third and the fourth. We will refer to this process as ELIM in this article referring to this straightforward process of elimination.

Entering a code is the best way to familiarize yourself with this interface. Try entering 1234 or any PIN of your choice below in Interaction 1. You can scan the QR code on your phone or click on the image or the link in the caption to access the interactive demo.

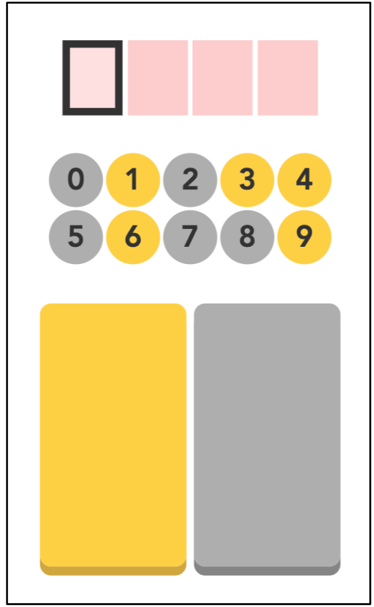 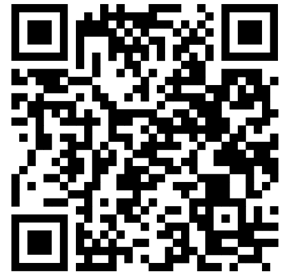

*Interaction 1: The IFTT-PIN interface with buttons of known colors.*
*Demo available at https://openvault.jgrizou.com/#/ui/demo_1x2.json*

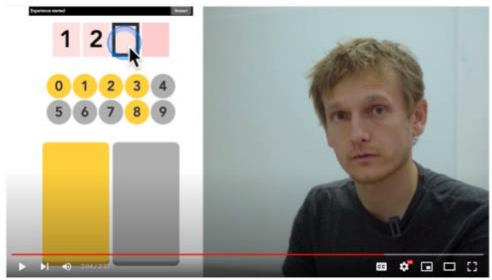 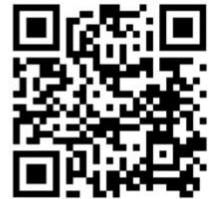

*Video 1: A step-by-step explanatory video of the IFTT-PIN interface with buttons of known colors.*
*Video available at https://youtu.be/DsqyD3eKX3E*

Congratulations! You now understand how to use the IFTT-PIN interface. We describe next what you can expect to learn from this article in more detail.

# Overview

In Interaction 1, you conveyed your intent to the machine via two big brightly colored buttons. The colors are meant to indicate what the buttons mean. It allows the machine to confidently translate button presses into their meanings, which seems to be a necessary condition for any interface to work.

 In what follows, we question this assumption and ask: Could a machine respond to user's commands without knowing exactly what those commands mean? Concretely, could the IFTT-PIN interface identify your PIN without knowing the colors of the buttons?

We call such an interface a self-calibrating interface because it does not need to be explicitly calibrated to each user's preferences. It rather learns those preferences on the fly. We will showcase and decipher such an interface using different versions of our PIN-entering machine.

The plan is as follow:

1. In section 1, we analyze the interface as presented in Interaction 1 and identify and name the various elements at play.

2. In section 2, we present the self-calibrating version of our interface. The buttons lose their pre-assigned colors. You get to decide the buttons' colors in your mind and use them as such. The machine will nonetheless figure out your PIN, as well as the color of each button. We will decipher how it works and refine our mental model to include what we learned.

3. In section 3, we scale our approach to continuous user's actions. A challenging task because it requires to train a machine learning classifier without access to known labels. To showcase this, our interface will no longer have buttons. You will place points on a 2D map and decide which areas are associated with which colors.

4. In section 4, we go deep into key implications of our approach that are sometime counter intuitive. For example, at no point our algorithm tries to classify user's actions into their meanings which allows the user's action-to-meaning mapping to change during the interaction. Although relying on clear illustration, this section might require a little background in machine learning and can be skipped.

5. In section 5, we demonstrate the use of richer interaction modalities. You will enter your PIN by drawing small sketches and using spoken commands. Once again, you will get to decide what sketches or sounds you want to use, and the machine will identify both your PIN and a classifier mapping your actions to their meanings.

6. In section 6, we present previous academic work on the subject, position this work into the human-computer interface and machine-learning landscape, highlight some remaining challenges, and ask ourselves how this work can be useful with concrete examples from brain-computer interfaces to the study of language formation.

7. In section 7, we go on a wild tangent and explore why applying principles of self-calibration in our personal and professional life might make us better citizens and leaders. Intriguing for sure, we hope you will give this section a chance.

8. Section 8 is for additional resources.

9. Section 9 is a list of all demos, videos, and figures for quick access.

You do not need to have a theoretical background in computer science, machine learning, or human-computer interfaces to understand most of the content in this article. There is purposefully no math or equation but interested readers will find relevant scientific literature referenced in section 6.1. In some sections, we use machine-learning specific terms to highlight important details. These terms will be explained but you can also safely skip these sections if not directly relevant to you.

# 1. Analysis of the PIN-entering interface

Successfully entering a PIN using Interaction 1 means that you understand the principle of interaction between you and our interface. If you have a background in computer science, you probably even know how to implement this at home. Nonetheless, it is important that we break down and name the various elements at play.

First, we should name three components driving the user behavior:

- The user's **intent** is what the user wants the machine to do. Here entering a specific PIN, one digit at a time.

- The user's **meaning** is what the user wants to say to the machine. Here it is either: "*My digit is yellow*" or "*My digit is grey*".

- The user's **action** is what the user does to express their meaning. In Interaction 1, the user's actions are to press either the left or the right button.

Actions, meanings, and intents are illustrated in Figure 1 below.

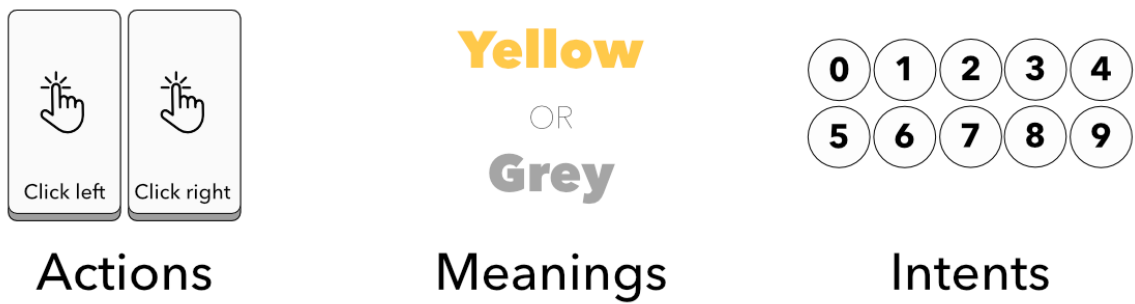

*Figure 4: Left: Actions the user can perform, either clicking the left or the right button. Middle: Meanings the user can convey, either Yellow or Grey. Right: Intents the user might try to achieve, either of the 10 digits.*

In short, an action conveys a meaning that is used to infer an intent. But inferring an intent from a meaning requires a bit of context. With our interface, the context is a user wanting to type a PIN via our interface. In other words, we assume that:

- The user aims to type a PIN one digit at a time. Thus, their current intent is to type one of ten possible digits.

- The user follows the established convention of indicating the color of the digit they want to type. Thus, the possible meanings are yellow or grey.

- The user can perform one of two actions, pressing either the left or the right button. This is constrained by the design of our interface.

- The **mapping between the user's actions and their meanings is known**. Pressing the left button conveys the meaning yellow and the right button conveys the meaning grey.

The latter is the most important element of our story; it makes explicit that there is a pre-existing shared understanding between the user and the machine about the meaning conveyed by the pressing of each button. This information is made salient by the colors displayed on each button, the left button is yellow, and the right button is grey. We call this information the action-to-meaning mapping (Figure 5).

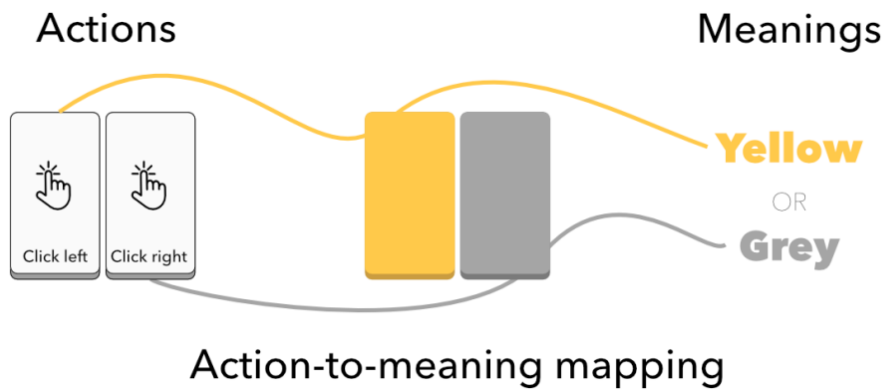

*Figure 5: Action-to-meaning mapping used in Interaction 1. The left button is associated with the meaning yellow, and the right button with the meaning grey.*

Knowing this mapping, the machine can reason as follows: "**If** a user presses the left button (*action*), **then** it indicates that their digit is currently yellow (*meaning*), **thus** their digit is among the yellow-colored digits and all the grey digits can be discarded (*intent*)."

This is the ELIM algorithm, see Figure 6 below and notice that the direction of inference is from left to right.

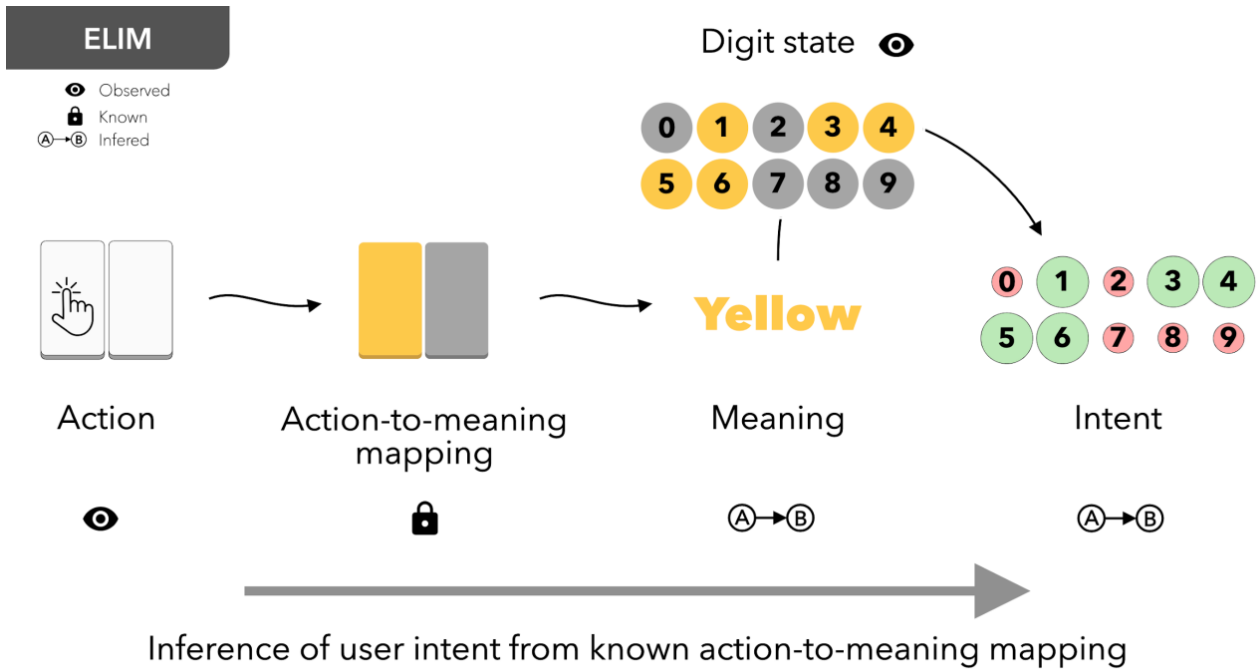

*Figure 6: The ELIM algorithm used in Interaction 1 . Each time the user clicks on a button, we can refer to the action-to-meaning mapping to eliminate all digits that are not the color meant by the user.*

Using ELIM, and by iteratively changing the color applied to each digit, we can narrow the possible digits down to the one the user has in mind (Figure 3).

To visualize this process while using the interface, we added a side dashboard to IFTT-PIN that displays the history of your clicks with respect to each digit. Because there are 10 possible digits, we are showing 10 individual panels, one for each digit (Figure 7).

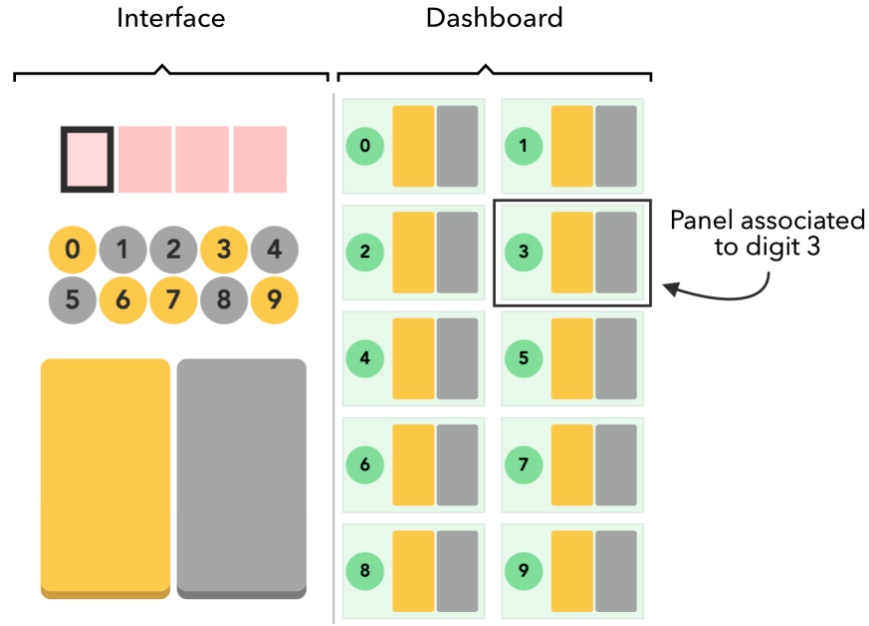

*Figure 7: IFTT-PIN with a side explanatory panel showing the inner working of the algorithm used for decision making. Each panel shows the history of interaction of the user with respect to a specific digit.*

After each click of the user, a dot is placed on the button that was pressed (left or right) to signify a click was made on that button. But this dot is colored differently for each panel. It is colored with the color that was assigned to the panel's associated digit when the user pressed the button.

Figure 8 shows this process for two digits, 0 and 1. The digit 0 is yellow, the digit 1 is grey, and the user clicks on the left, yellow, button. A dot is placed on the left, yellow, button on both the panels associated with both digit 0 and 1. But this dot is colored in yellow for the digit 0 and in grey for the digit 1. The digit 1 can be discarded because a grey dot is placed on the yellow, left, button. The logical being that, if the user wanted to type a 1, they would have used the grey, right, button instead, because the digit 1 was grey. The digit 0, however, remain a valid possibility.

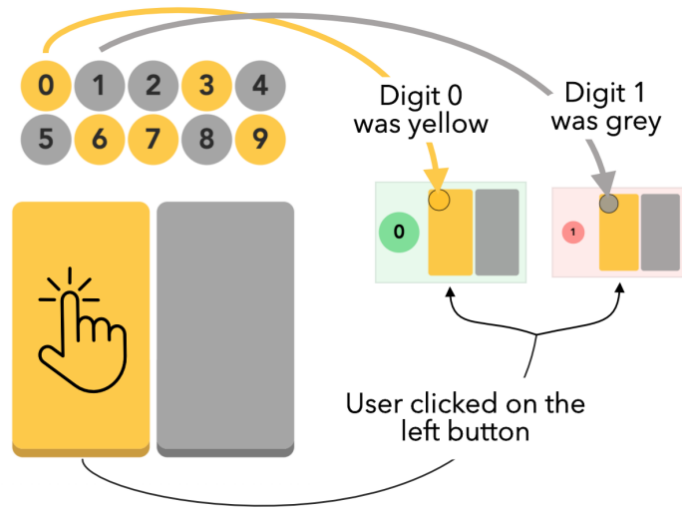

*Figure 8: How the side panel works. For each side panel, a dot is placed on the button clicked by the users (left button), and the color of this dot is the color of panels' associated digit at the time the button was pressed (yellow for 0 and grey for 1).*

The valid/invalid status of each digit is shown visually by the color and size of each panel. When a digit is still valid, its panel is green and large. When a digit is discarded, the panel is red and smaller (Figure 8).

Try typing a PIN on the explanatory interface below while monitoring the elements on the dashboard. Make sure to understand how to interpret this side dashboard, we will use it in various form all along this article. A video is also available for an interactive walkthrough.

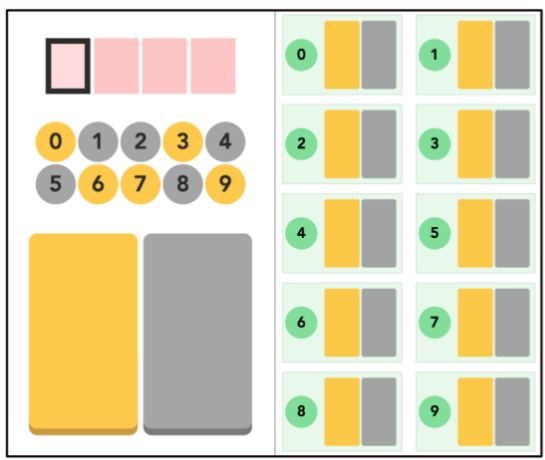
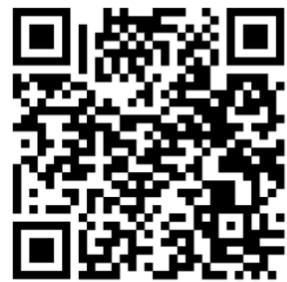

*Explanation 1: The IFTT-PIN interface with buttons of known colors and a side explanatory panel. Demo available at https://openvault.jgrizou.com/#/ui/tuto_1x2.json*

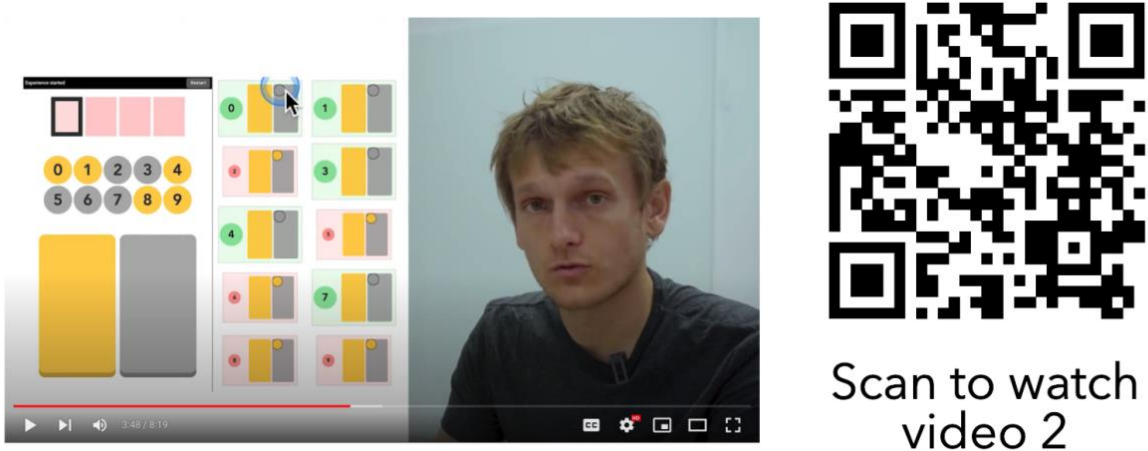

*Video 2: A step-by-step explanatory video of the IFTT-PIN interface with buttons of known colors and a side explanatory panel. Video available at https://youtu.be/xP3qJ1V28Ws*

You should now have a clear understanding of how our PIN-entering interface works and have acquired some element of language such as action, meaning, intent, context, and action-to-meaning mapping.

In the next section, we introduce the self-calibrating version of this interface in which buttons have no pre-defined colors. You will get to decide the colors of each button in your mind and never explicitly tell the machine about it. The machine will nonetheless be able to identify your PIN and the colors of the buttons.

# 2. Self-calibrating PIN-entering interface

What if the buttons had no colors? In other words, what if the action-to-meaning mapping - between the position of the buttons (left/right) and their meaning (yellow/grey) - was not pre-defined?

If we look back at Figure 6, the chain of inference is now broken. Without a known action-to-meaning mapping, we cannot infer what the user means, thus we cannot infer the user intent.

To solve this problem, the usual approach (which we don't want to use in this article) is to first learn the action-to-meaning mapping that our user would like to use. The aim is to calibrate the interface to the user preference before they start using it. To do so, the user is asked to follow a calibration protocol, which can be direct or indirect.

When direct, we simply ask the user for elements of the action-to-meaning mapping using another, already calibrated, interface. For example, we give each user a digital paint brush that can take two colors (yellow or grey) and ask them to color each button in the way they would like to use them (Figure 9).

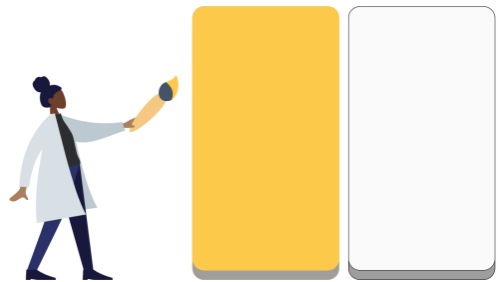

*Figure 9: Illustration of direct calibration of an interface. The user is given a yellow paint brush and can decide which button to paint in yellow, indicating to the machine their preferred action-to-meaning mapping.*

When indirect, the calibration procedure directly uses the interface, but the user is asked to achieve a specific, known, goal. For example, we ask the user to type the digit 1. Knowing the digit, we can reverse the inference pipeline and follow this reasoning: "Knowing that the user is typing a 1, if the digit 1 is yellow and the user is pressing the left button, then the left button means yellow" (respectively for all buttons and colors). Figure 10 illustrates this reasoning.

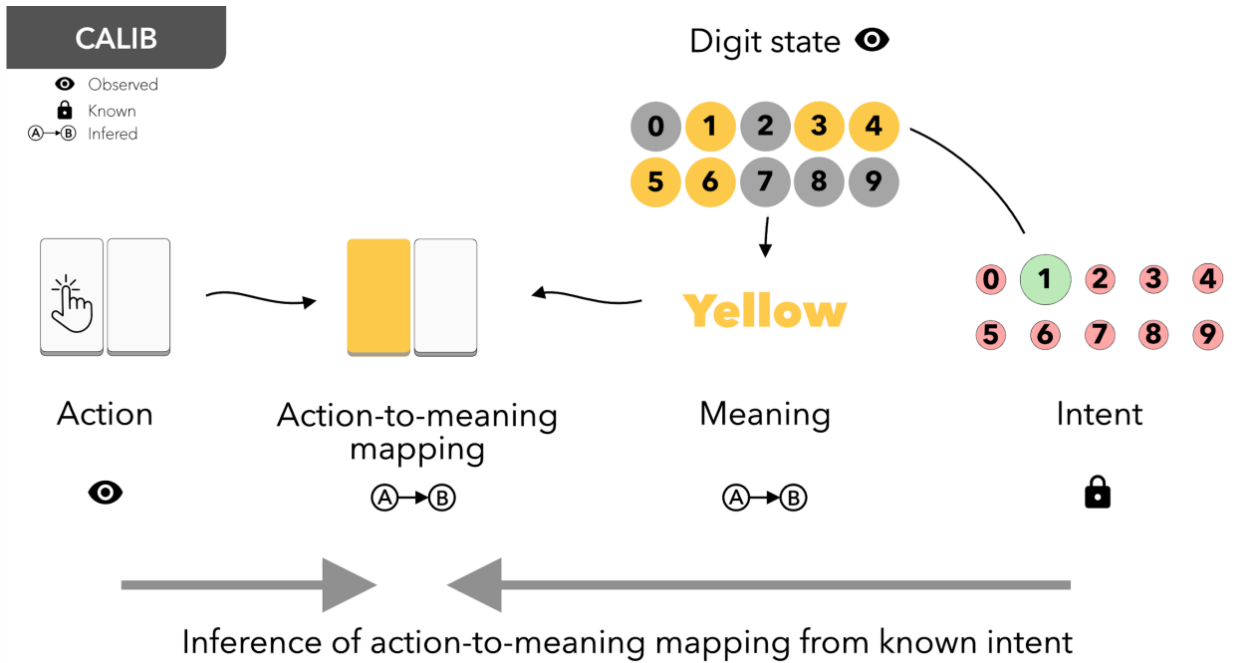

*Figure 10: A typical calibration procedure, by enforcing the user to achieve a specific intent, we can infer the action-to-meaning mapping by observing the user's actions (button clicks) given the current state of the system (digits' colors).*

We refer to this process by the name of CALIB, because it is the most common method to calibrate an interface to user preferences.

But CALIB is exactly what we do **not** want to resort to here. We want to investigate if and how our interface could self-calibrate. We want to explore if and how one can identify the user intent without knowing the action-to-meaning mapping. Why? For intellectual curiosity more than practicality. We will review potential applications in section 6.3 but for now let's not ask ourselves why and focus on the how.

Let's thus imagine that a user is arbitrarily assigning colors to each button in its mind and uses the interface that way - without telling the machine about its color choice, nor its intended PIN.

The machine is in trouble, it does not know what digit the user wants to enter, and it does not know what the user means when pressing buttons. The ELIM reasoning used in section 1 collapses because we cannot follow the logical path: "**If** the user presses the left button, **then** they mean that their digit is currently yellow".

If anything, this line of reasoning turns into: "**If** the user presses the left button, **then** their digit is either yellow or grey with equal probability, **thus** I cannot make any decision." That sounds like a dead end.

Before explaining how we solve this problem, we think you should see its results in action and experience how it feels to be able to arbitrarily choose buttons' colors.

Interaction 2 works the same way as in Interaction 1 but no colors are displayed on the buttons. You choose the colors in your mind. To make it more interesting, we increased the number of buttons from 2 to 9 (Figure 11). That way, instead of having 2 possible ways to assign colors to the buttons, you now have 510 ways[2].

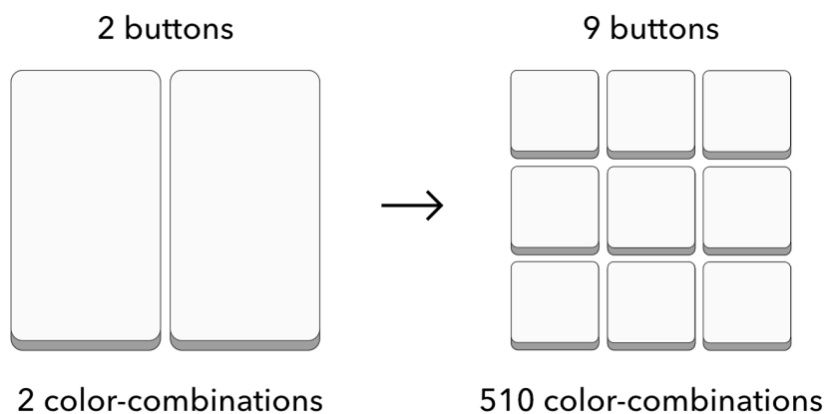

Figure 11: We increase the number of buttons from 2 to 9 to make the color choice more interesting. Buttons can only be yellow or grey and we have now 510 possible combinations instead of 2.

The colors are only in your mind, and you can assign them as you please. For example, in Figure 12, we show how three users decided to assign colors on the buttons. Providing that there is at least one button for each color and that you stick with the same color pattern during the interaction, the machine will infer both your PIN and the colors of the buttons.

---

[2] We need at least one button for each color. If all buttons are yellow and your digit is grey, you would not be able to express the meaning grey. The number of combinations is 2^N - 2, with N the number of buttons. And -2 because there are 2 invalid combinations, all yellow or all grey.

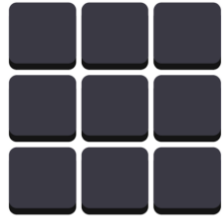

Buttons have no colors to start, they are all black.

Buttons can be used differently by each user

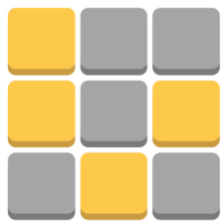

User 1 preference

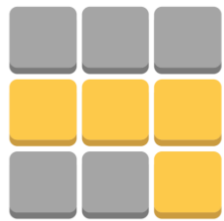

User 2 preference

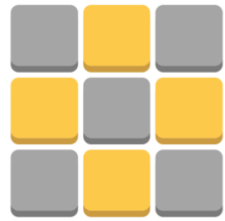

User 3 preference

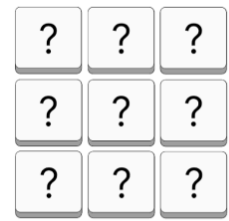

What color pattern will you use?

*Figure 12: Example of how different users might choose to assign colors to each button. You are free to choose your own combination, and you need at least one yellow and one grey button.*

Try to use this interface multiple times, entering different PINs like you did in Interaction 1 and using different color patterns. You do not have to use all buttons every time. Only the button you used will be identified and colored in by the interface. The others will remain black until you use them[3].

---

[3] If you do not use some buttons, the machine does not have any information about them, and has no way to infer their colors.

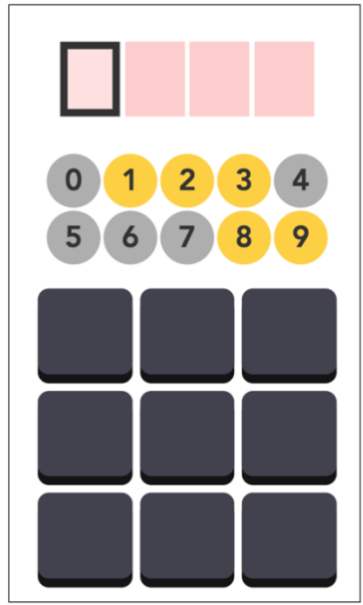
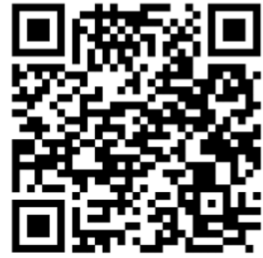

*Interaction 2: The self-calibrating version of IFTT-PIN with uncolored buttons.
Demo available at https://openvault.jgrizou.com/#/ui/demo_3x3.json*

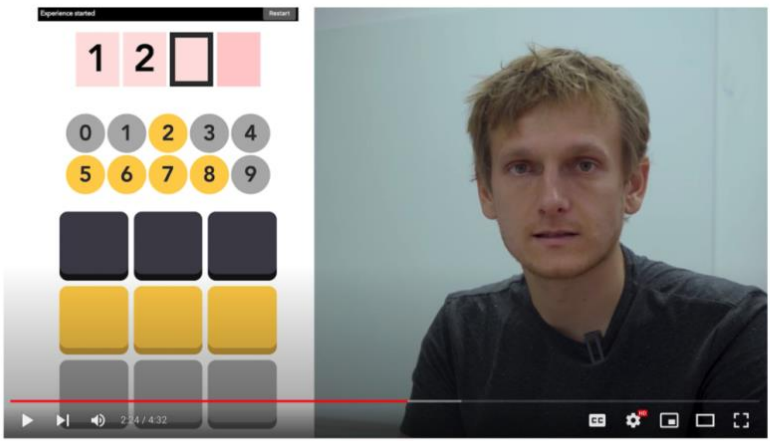
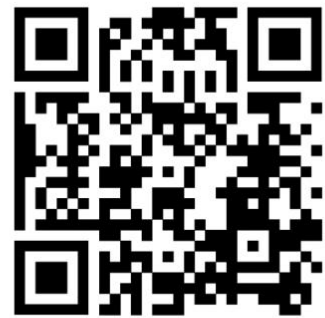

*Video 3: A step-by-step explanatory video of the self-calibrating version of IFTT-PIN with uncolored buttons.
Video available at https://youtu.be/upKejh4ZgUc*

It is an interesting feeling, isn't it? We are not used to having this level of choice when using the machines around us. To understand how this works, we shall look at the problem from a new angle.

In section 1, we defined the following components: intent, meaning and action. We understood that an action conveys a meaning that can be used to infer an intent. And we have seen that this logical path requires a context that allows to deduce meanings from actions and intents from meanings.

This context is the list of assumptions embedded within the interactive process it-self. We assume that users want to type one of ten possible digits. We assume that they indicate the color of the digit they have in mind. And we assume that they press buttons to send their feedback. All these assumptions remain, but one can be added which was hiding in plain sight.

We assumed all along that a button can have one and only one meaning - yellow or grey - never none and never both. This assumption was hard to formulate before because colors were visibly assigned to each button, it was too obvious to be noticed. The assumption that one button equals one meaning is so ingrained in our interaction with machines that we sometimes forget it is part of the convention.

Why can this help us? Because it is something we can observe. By measuring breaches of the "yellow or grey" assumption, we can solve the self-calibration problem.

How can we measure such breaches? By making hypotheses. Because we know the user is trying to type one of the ten possible digits, we can imagine ten different worlds, each with the user trying to type one specific digit. One hypothetical world for each of the ten digits. In each of these worlds, because we hypothetically enforce the digit the user is trying to type, we can easily infer the colors of the buttons using the same reasoning as the CALIB algorithm: "**If** the user is trying to type a 1 (intent), **then** each time the user presses a button (action), we can assign the current color (meaning) of the digit 1 to that button".

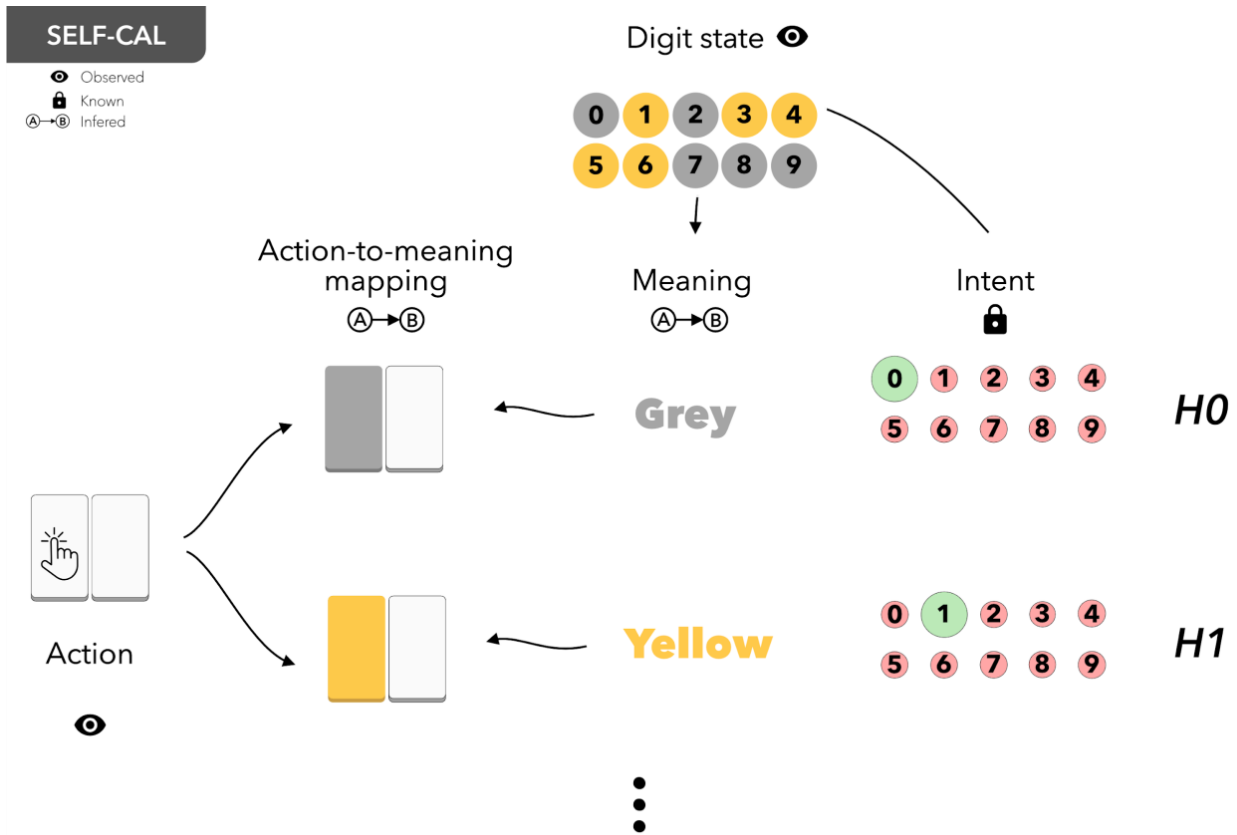

*Figure 13: The SELF-CAL algorithm reuse the same principle as the CALIB algorithm but applies it to all possible user intents. These hypothetical intents are shown on the right side and marked H0, H1, etc.*

In essence, we are performing ten CALIB procedures in parallel, one for each digit. In other words, we are building ten different action-to-meaning mappings, one for each digit.

But because the user is entering only one of the ten possible digits, only one of the button-to-color maps will be valid. Only one will conform with the "yellow or grey" assumptions. For all other hypotheses, at some point during the interaction, it will look as if the user was trying to press some buttons to mean **both** yellow and grey - signaling a breach of our "yellow or grey" assumption, which is enough to discard the associated digit.

In other words, when, from the point of view of a given digit, the same button has been used to mean both yellow and grey, then that digit cannot be the one the user has in mind because it is incompatible with our assumption that one button has one and only meaning.

We name this process SELF-CAL. To visualize it while you enter a code, we added a side dashboard acting similarly to the one in section 1. There are ten panels, one for each digit (Figure 14). In each panel, the buttons are shown and will be populated with dots after each click of the user. Each dot will be colored differently for each panel using the color that was assigned to the associated digit when the user pressed the button.

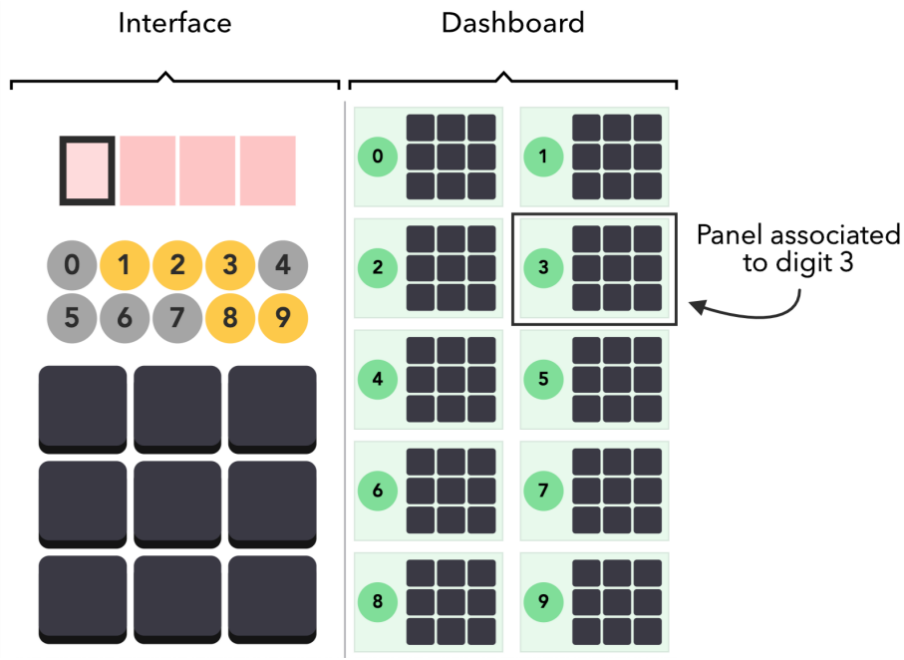

Figure 14: The self-calibration IFTT-PIN interface with a side explanatory panel.

This time, because the buttons have no color, instead of comparing the dot color with the button color, we compare the dots on each button between themselves. If all dots on the same button are of the same color, the hypothesis is still valid. However, a button that is populated with dots that are both yellow and grey signifies a breach of the "yellow or grey" assumption and the hypothesis can be discarded.

Figure 15 shows the result of this process for digits 0, 1, 2, and 3 after a few clicks of the user in a particular run. The user was trying to type a 1, only used two out of the nine buttons and had done a total of three clicks so far.

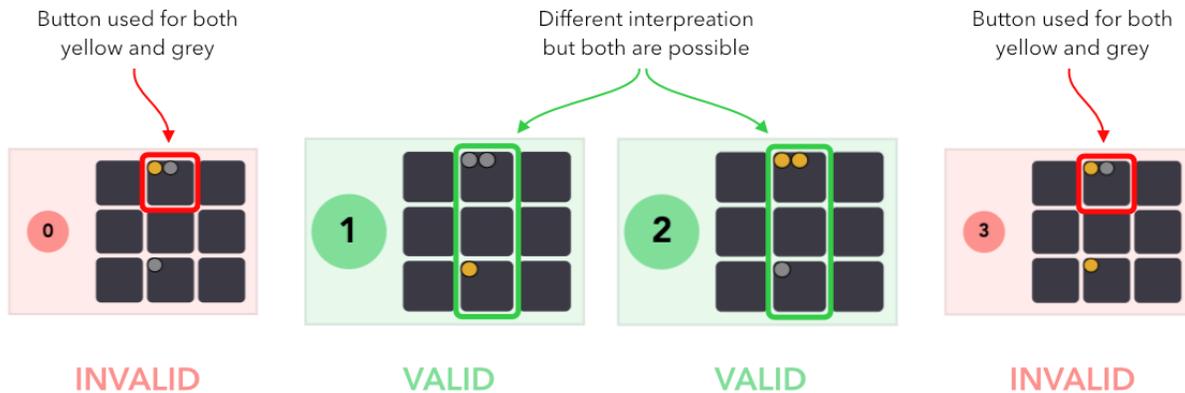

*Figure 15: Example of a subset of side panels state after 3 clicks from the user. The user is trying to type a 1 and sed only two buttons so far. Digits 0 and 3 are not valid because, if the user was trying to enter 0 or 3, they would have been using the middle-top button to mean both yellow and grey. Digits 1 and 2 are still valid but offer two opposite views of the user's action-to-meaning mapping, both conforming with the "yellow or grey" assumption.*

There are two things to notice on Figure 15. First, the digit 0 and 3 have already been discarded. This is because, if the user was trying to type a 0 or a 3, then they used the top button to mean first yellow, then grey (as indicated by the yellow and grey dot in that button for both hypotheses). But one button can only be used for one color, so the user is not trying to type a 0 nor a 3. Second, the digits 1 and 2 are both still valid despite having differently color of dots in each button. If the user was trying to type the digit 1, then they used the top button to mean grey and the bottom one the mean yellow. Reversely, if the user was trying to type the digit 2, then they used the top button to mean yellow and the bottom one to mean grey. Importantly, both options are still viable. The machine cannot decide yet which one is valid. Those two interpretations of the user actions are still possible, and more information is needed to pull them apart.

After a few more iterations, only one hypothesis will remain valid and free from any "yellow and grey" conflicts. At this point, the interface can be confident that the associated digit is the one the user wants to type. Now knowing the user intended digit, we also immediately know the button-to-color mapping the user had in mind all along. We are finding both what the user is trying to do and how they are trying to do it. We are self-calibrating.

You can interactively visualize this process directly on Explanation 2 below.

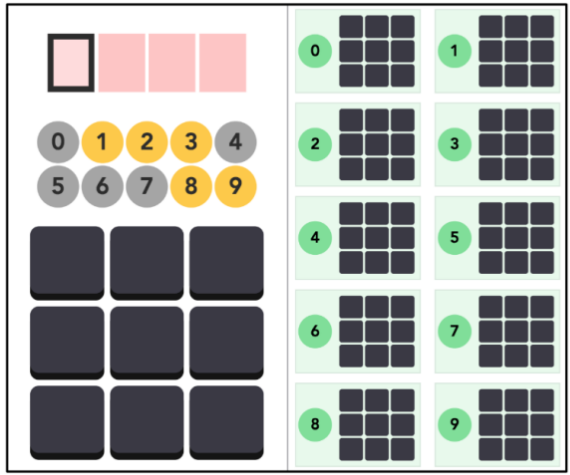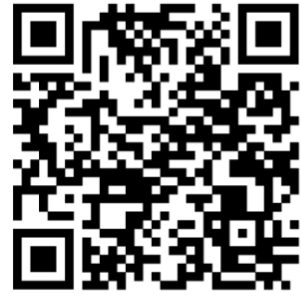

*Explanation 2: The self-calibrating version of IFTT-PIN with uncolored buttons and a side explanatory panel. Demo available at https://openvault.jgrizou.com/#/ui/tuto_3x3.json*

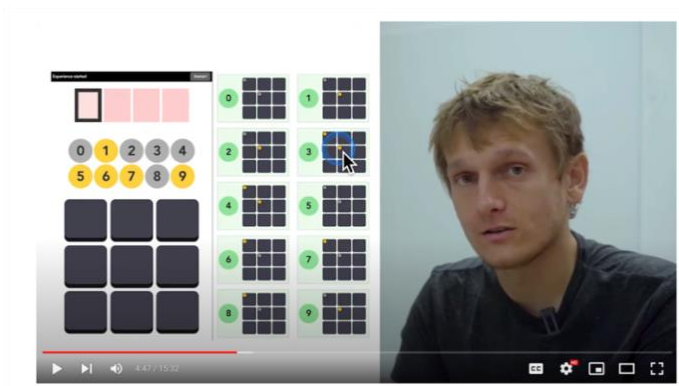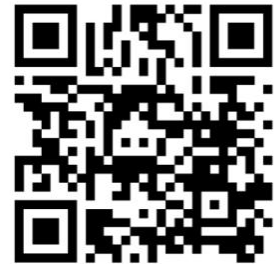

*Video 4: A step-by-step explanatory video of the self-calibrating version of IFTT-PIN with uncolored buttons and a side explanatory panel. Video available at https://youtu.be/OMlQRy_ZKFs*

After playing with the interface for a bit, you should have gained an important insight about the self-calibration problem. To solve it, we no longer try to understand what the user means when pressing buttons, we simply gauge for which digits the actions of the user remain consistent in time.

It logically takes more time to identify the first digit when we do not know the color of each button than when we know them. And it is interesting to observe how alternative interpretations of the same user's actions remain valid quite far into the identification process. As an exercise, you can try pushing SELF-CAL to its limit by finding a click strategy that never allows the interface to identify the digit you want to type. For example, focus on only two digits and try to keep their associated panel always valid by carefully choosing the button you press depending on the current color of those two digits. Succeeding would show you truly understand what is going on. If curious, we show one such strategy in section 4.3.

Before ending this section, we need to focus on what happens once a first digit is identified. At the exact moment the machine identifies the first digit, we are essentially back to the CALIB approach. We know what digit the user was trying to enter, so we can infer the color of each button the user pressed. The interface is thus capable of displaying the right colors, yellow or grey, to all the buttons you used.

Notice how the button's colors are also shown in each panel of the side dashboard. It becomes a common prior information about the button-to-color mapping that can be used to identify the next digit. Next time you press one of these buttons, the interface already knows what you mean and can directly reuse the ELIM reasoning from section 1. For all other buttons, we can keep using the SELF-CAL approach, looking for breach of the "yellow or grey" assumption.

In a strange twist, this implies that the reasoning behind ELIM is equivalent to the reasoning behind SELF-CAL when all hypotheses agree on the button's colors. ELIM is only a particular case of SELF-CAL in cases where prior information is available.

We can reframe thus the ELIM inference process as follows: "**If** the user is trying to type a 1, and *if* the color of the button the user is pressing is different from the color applied on digit 1, *then* the same button is being used to express two different colors. **Thus,** the user is not trying to enter the digit 1. **Else** they might be typing a 1". Convoluted but strictly equivalent and a powerful way to reframe human-machine interaction scenarios that enabled us to exploit a hidden "yellow or grey" assumption to solve the self-calibration challenge.

The remainder of this article considers how to scale this "yellow or grey" logic to continuous user's actions. In the next section, you will discover a version of our interface with no button. Instead, you will place points on a 2D map, and you will get to decide which areas are associated with which color.

# 3. Can this approach scale to continuous signals?

Up to now, we considered discrete button presses and our logic was based on identifying if the user was using the **same** button (action) for different colors (meaning). This notion of "**same**" was easily measurable with discrete button events. But when the user's actions are more complex, such as drawings, sounds, gestures, brain signals, or nerve impulses, an action will never be represented twice in the same way. We call these continuous signals.

When dealing with continuous signals, we can no longer define a notion of "same" ahead of time. It must be learned from the user data.

To explain this problem, we designed an interface with no button. Instead of pressing buttons, you will place points on a map. The points can be placed in a yellow area to mean "My digit is yellow", or in a grey area to mean "My digit is grey" as explained in Figure 16.

Because we are in a self-calibrating scenario, the color map is not defined in advance nor displayed on the interface. It is to be defined by you and resides in your mind. You decide which areas of the maps are yellow or grey and the machine must figure out both the map you use and your PIN.

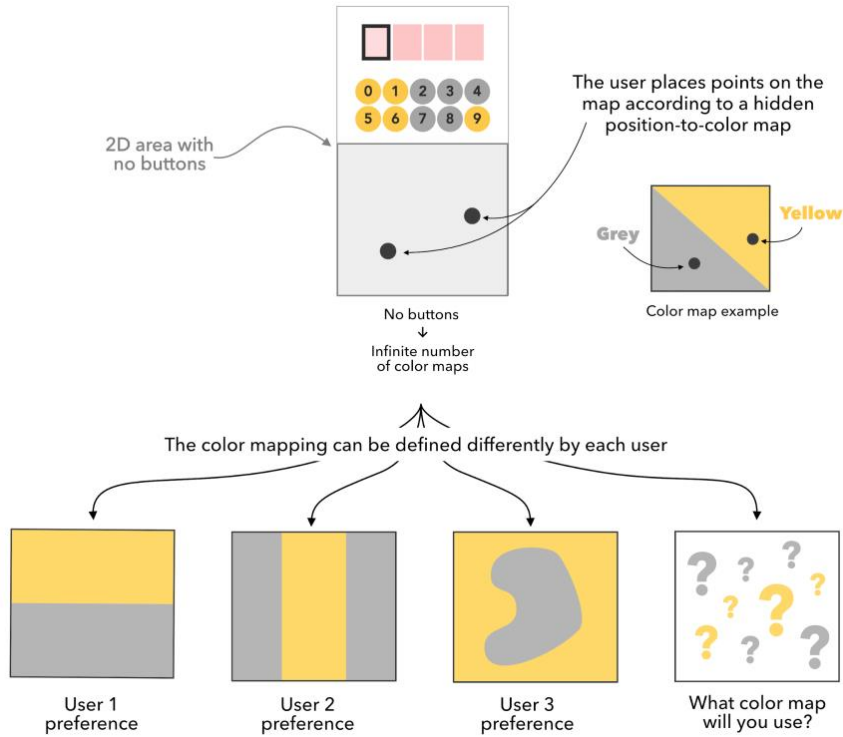

*Figure 16: Explanation of the continuous color-to-meaning mapping. Instead of pressing buttons, user now click on a 2D location inside an area identified as yellow or grey as per the user's hidden color map preference.*

We think it is best to try this new version of IFTT-PIN before explaining how we solve it. You can try it on Interaction 3 below.

Before you start, remember this is a hard problem and be patient. It might take between 10 to 20 clicks for the machine to identify your first digit.

Start simple, for example typing the code 1234 by assuming the left part of the map is yellow and the right part is grey, in which case the result would look like Figure 17. We recommend watching the associated video if unsure about what to do.

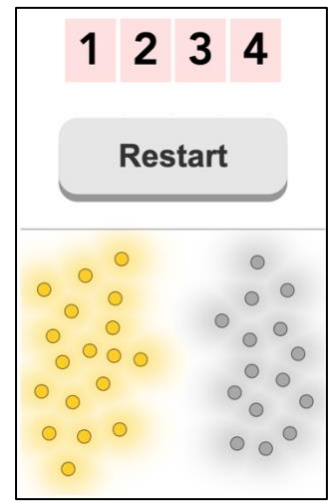

*Figure 17: Helper to try the touch version of IFTT-PIN*

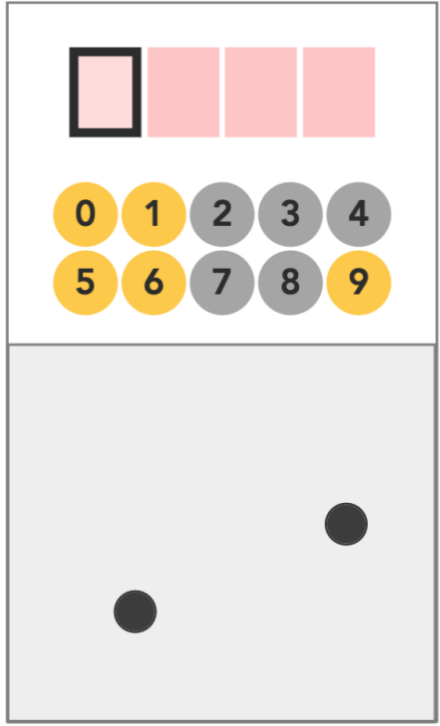 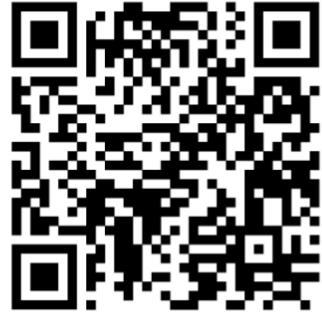

*Interaction 3: The self-calibrating version of IFTT-PIN with no button.*
*The user places points on the 2D map and decides which areas on the map are yellow and grey in their mind.*
*Demo available at https://openvault.jgrizou.com/#/ui/demo_touch.json*

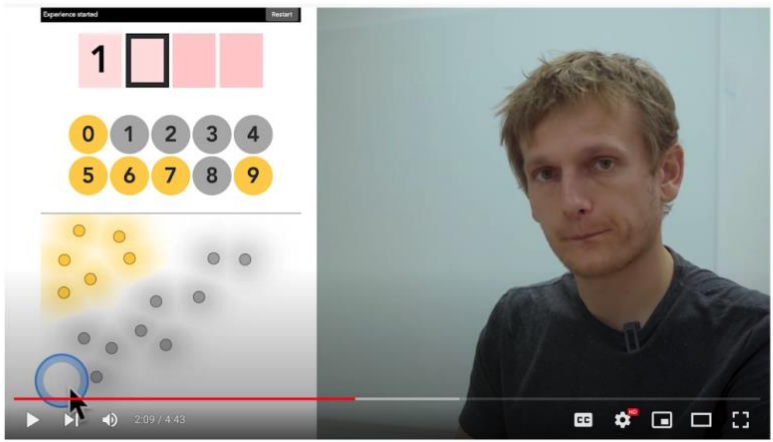 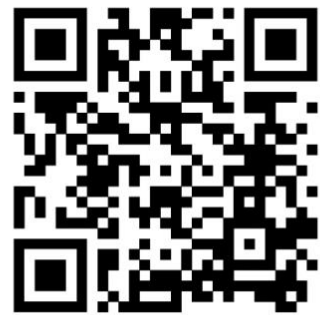

*Video 5: A step-by-step explanatory video of the self-calibrating version of IFTT-PIN with no button.*
*Video available at https://youtu.be/b4NjrMB6VLs*

Points placed on the map are an example of continuous signals. You never clicked twice exactly in the same place. This means we cannot tell if two points represent the same color just by looking at them, even more so in the beginning when no structure has emerged from the data. Ask a friend to guess what you are doing; it is likely they will be clueless. So how are we solving that problem?

To ground our explanation, we first need to squeeze the concepts covered so far into one word: **consistency**. This will help our brains navigate this chapter.

From section 1 and 2, both ELIM and SELF-CAL algorithms can be described as measuring the consistency of the user while using our interface. In both cases, we have been detecting breaches of consistency:

- In section 1, we defined consistency as: _clicking on a button of the same color as the digit we want to type_. And a breach of consistency was looking for digits that were not of the same color as the button clicked by the user.

- In section 2, we defined consistency as: _using a button to only mean one color - the "yellow or grey" assumption_. And a breach of consistency was looking for digits which, if a user was entering them, that user would have been pressing the same button to mean both yellow and grey.

To scale our approach to continuous signals, we need to define a consistency metric for continuous signals. Previously we defined consistency with statements like "a button of the **same** color" and "only mean one color - yellow **or** grey". But the notion of "same" and "or" are no more applicable as all signals are different now. We need a looser measure of similarity between signals.

While we cannot be in the mind of every person using this interface, we can nonetheless come up with broad principles of how most people should behave when deciding how to allocate colors and place points on the 2D map.

For example, we can assume that users will define yellow and grey areas that are easy to differentiate, so they could remember where to place a yellow or a grey point when required. Another common assumption is that the user will place points of the same color "close" to each other, where closeness could be measured by the Euclidean distance between two points.

Summarizing these assumptions, we can define consistency for continuous signals as: _using a simple color map_. Where simple is defined by the ability to easily differentiate between the yellow and grey points. Figure 18 shows examples of simple and complex color maps.

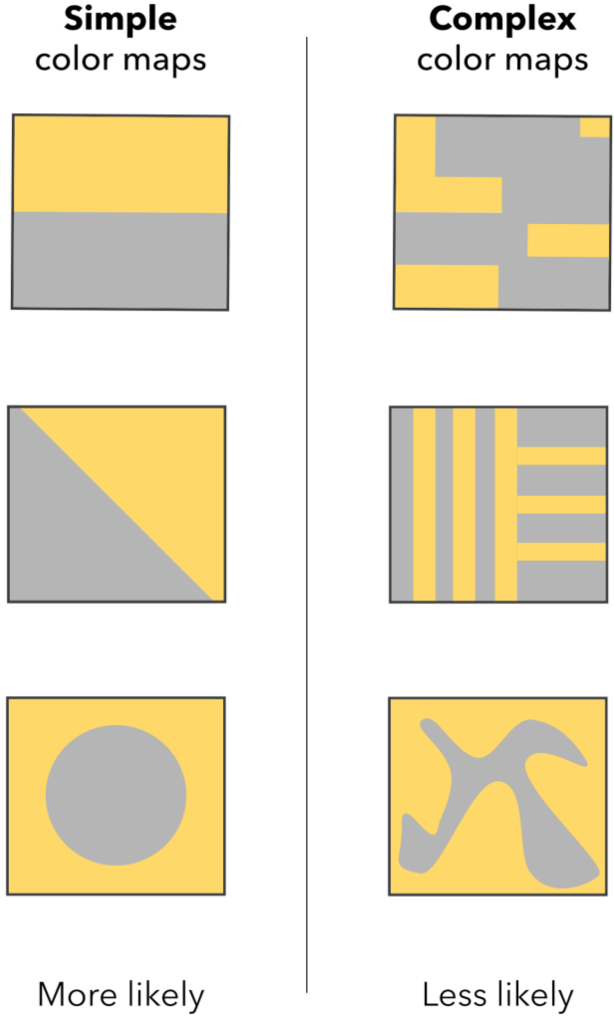

*Figure 18: Example of possible color maps used by a user. Given our consistency assumption, our metrics assumed that maps on the left side are more likely to be used by a user than maps on the right side.*

If consistency is defined by simplicity and the ability to easily differentiate between the yellow and grey points, the next question is "how can we measure simple"?

## 3.1. How can we measure "simple"?

This notion of a "simple color map" is very loose. We need to be able to put a number on this notion for our algorithm to work.

Luckily, machine learning experts invented classifiers. Given a set of colored points (a training set), a classifier can extrapolate and generate a color map that "best explains" the training set. Several assumptions are made by machine learning practitioners to define this "best explains" criteria.

A common assumption is that the simpler the map the better[4]. This assumption is often included as a regularization term in the classifier's cost function that penalizes solutions that lead to complicated maps. Complicated maps are usually defined as non-smooth maps with many sharp changes at their frontiers or that form a lot of isolated islands. Creating such complicated maps usually requires large weights/parameters value in the decision function, which the regularization terms penalize.

How elegant. Machine learning researchers found a way to embed notions of "simplicity" and "best explain" in quantitative terms. Which is exactly what we need to measure consistency in our scenario. Because these assumptions are baked into classification algorithms, they are the perfect tools to measure the consistency of our user when dealing with continuous signals. More precisely, the prediction accuracy of a classifier trained on data generated by our users can be seen as a direct measure of their consistency. If we can train a good classifier, then the underlying mapping is "simple", and the user is consistent. If we cannot train a good classifier, then the underlying map is judged too complex. Regularization terms prevented the algorithm from fitting a convoluted decision function which indicates that the user is inconsistent.

It is no surprise that classifiers are perfectly matching with our problem. The assumptions used when designing machine learning algorithms are made by and for humans trying to make sense of the world. They are meant to reflect the way we, humans, generate and classify things in the world. Hence, similar assumptions emerge when we think of how a user will generate and use a color map for our PIN interface. These assumptions are not always true, but they are the best we can do without more explicit prior information.

---

[4] See Occam's razor principle and the bias–variance tradeoff in machine learning.

## 3.2. How can we leverage classifiers to solve the self-calibration problem?

Classifiers suit our needs in theory but using them in practice to solve our self-calibration problem requires some inventivity. Indeed, a classifier needs to be trained on labelled data but, because we are in a self-calibration scenario, we do not have access to such labels. In other words, we need to know the colors (meanings) of the points (actions) generated by a user to be able to measure the user consistency, but we do not have access to the point-to-color (action-to-meaning) mapping.

To get around this problem, we can use the SELF-CAL trick explained in section 2. Because we know the user is entering one of ten digits, we can generate ten different datasets, each with the same data points but with different labels/colors for each hypothesized digit.

To identify the intended digit, we then compare the consistency of each hypothetical labelling of the data. As explained above, a good proxy for consistency is the quality of a classifier trained on the data. We use a SVM classifier with a RBF kernel and measure the cross-validation classification accuracy for this work, but other classifiers and/or metrics can be imagined within the broad spectrum of machine learning tools.

To explain differently, we use the process of training a classifier as our consistency filter. If we can train a good classifier on the data, it means that the assumptions baked into classification algorithms are respected. Hence the map used by the user can be considered as simple, and therefore the user is consistent. However, if we cannot train a good classifier, then some regularization terms stand on our way and the map would need to be too complex to account for all observations. Hence the map used by the user is considered too complex and therefore the user is considered inconsistent. See Figure 19 for a visual example.

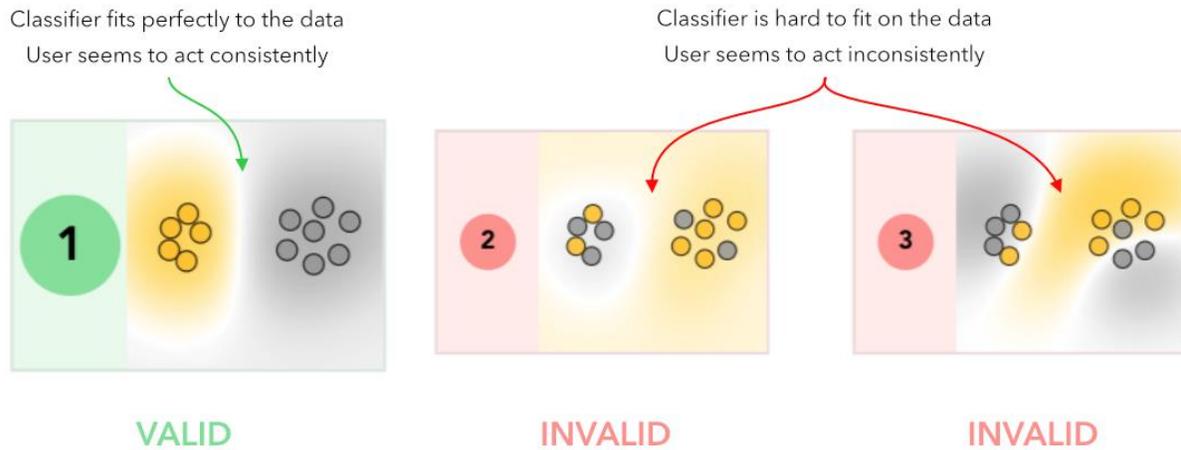

*Figure 19: Examples of different labeling of the use data according to digit 1, 2, and 3. For each labeling, a independent classifier is trained and the resulting color-map is displayed in the background. Visually, the color map on the left associated to digit 1 is simpler than the other two. This will be reflected in the classification accuracy metrics of the classifier and tells us that the user is more likely to be entering digit 1.*

The last challenge is to decide when to stop and pick a digit. We need to decide when a labelling system (i.e., a digit) is significantly more consistent than all others. Because we deal with continuous signals, we can never be 100% sure that we have the right answer. But we can nonetheless run statistical tests and agree on a threshold for which we are happy to claim that one hypothesis is statistically more consistent than all the others. In the beginning, when only a few data points are observed, all hypotheses will remain valid as it will be easy to train a good classifier for each hypothesis. But as more data is collected, one hypothesis will stand out as significantly more consistent than all others. The digit associated with this dataset should be the digit the user is trying to enter.

The all process is easy to understand visually using our side panel approach as seen in explanatory interface 3 below. Try it and notice how each hypothesis assigns different labels to your actions. As a result, each hypothesis builds a different classifier, resulting in a different color map, to explain your actions. After enough clicks, it becomes obvious which digit you are typing because all others hypotheses lead to more complex maps - indicating a breach in our definition of consistency for continuous signals: using a simple color map.

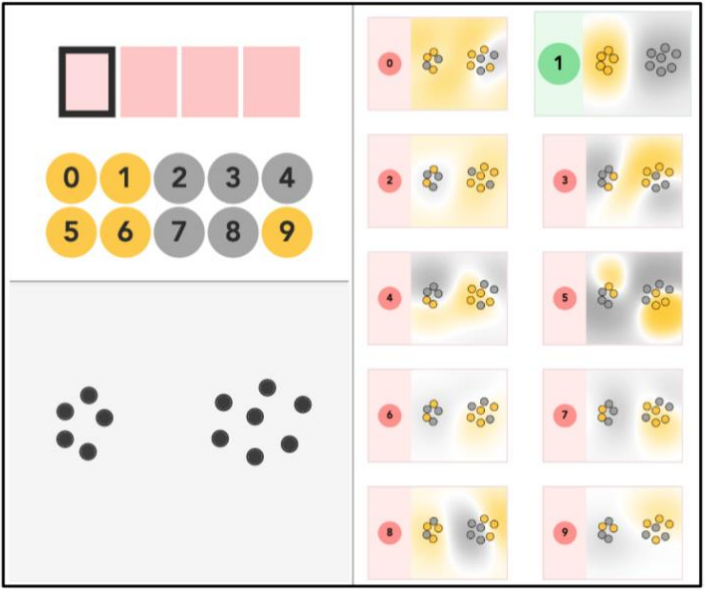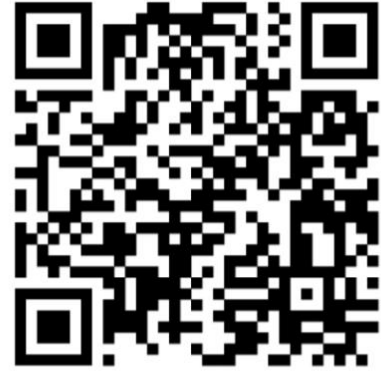

*Explanation 3: The self-calibrating version of IFTT-PIN with no button and a side explanatory panel. Demo available at [https://openvault.jgrizou.com/#/ui/tuto_touch.json](https://openvault.jgrizou.com/#/ui/tuto_touch.json)*

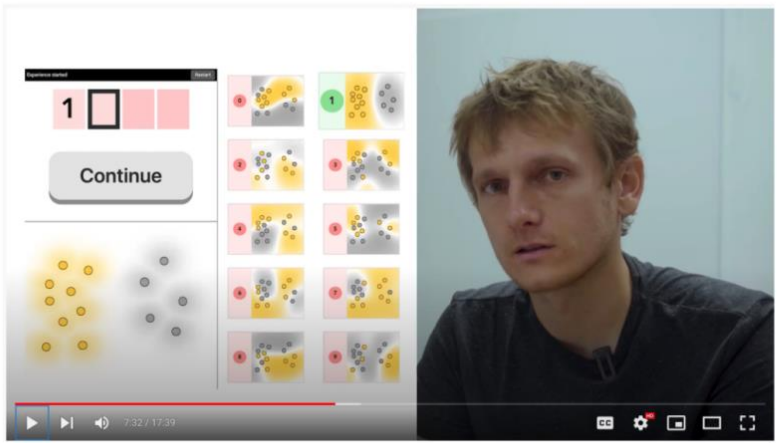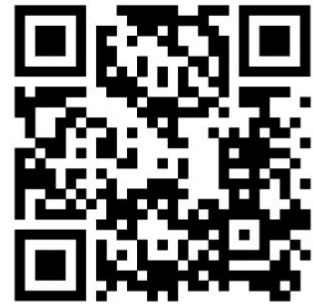

*Video 6: A step-by-step explanatory video of the self-calibrating version of IFTT-PIN with no button and a side explanatory panel. Video available at [https://youtu.be/ZUI7zbScUTk](https://youtu.be/ZUI7zbScUTk)*

I personally find this demo the most compelling in this article. Trying to challenge it with complex color maps, or trying to force false prediction, is a good exercise to verify that you understand the algorithm. In the next section, we dig deep and expose some surprising, if not counter intuitive, implications of using SELF-CAL.

# 4. Important implications

There are important implications that machine-learning connoisseurs should understand at this point:

1. We transform an unsupervised learning problem into a supervised learning problem, which allows handling unstructured and deceptive datasets.

2. We do not seek to classify users' actions into their meanings; thus, we allow our model of the user's action-to-meaning mapping to change during the interaction. As more data is received from the user, our method converges to the performance of a fully calibrated system.

3. Our stopping criteria is both data and task dependent. At first, understanding the data is the bottleneck, but once enough data is received from the user, solving the task becomes the bottleneck. SELF-CAL embeds both constraints into one stopping criteria and allows a smooth transition between the two limit cases - from having no prior to having a known, fixed, classifier.

We address these points and their implications in detail below.

## 4.1. We transform an unsupervised learning problem into a supervised learning problem

[Unsupervised clustering algorithms](#) are designed to identify groups of points that are similar to each other when no labels are available.

The notion of group and of similarity are based on assumptions about the process generating the data. For example, in our PIN-entering interface, we could assume that the user will generate points of the same color close to each other and that yellow and grey points will form two well separated clusters. In other words that data are generated from two "non-overlapping" Gaussian distributions, one for yellow points, one for grey points.

Starting from such assumptions, an instinctive approach to solve the self-calibration problem would be to first find those clusters in the data. Then assign colors to each cluster, for example using a rule of thumb based on the task (e.g., label proportion - if some colors are known to be used more frequently) or by considering all possible combinations. And finally, replay the history of interactions knowing the colors for each point to identify the intended digit.

Figure 20 compares this unsupervised approach (UNSUP) with the self-calibration method (SELF-CAL) on data organized in two well separated clusters.

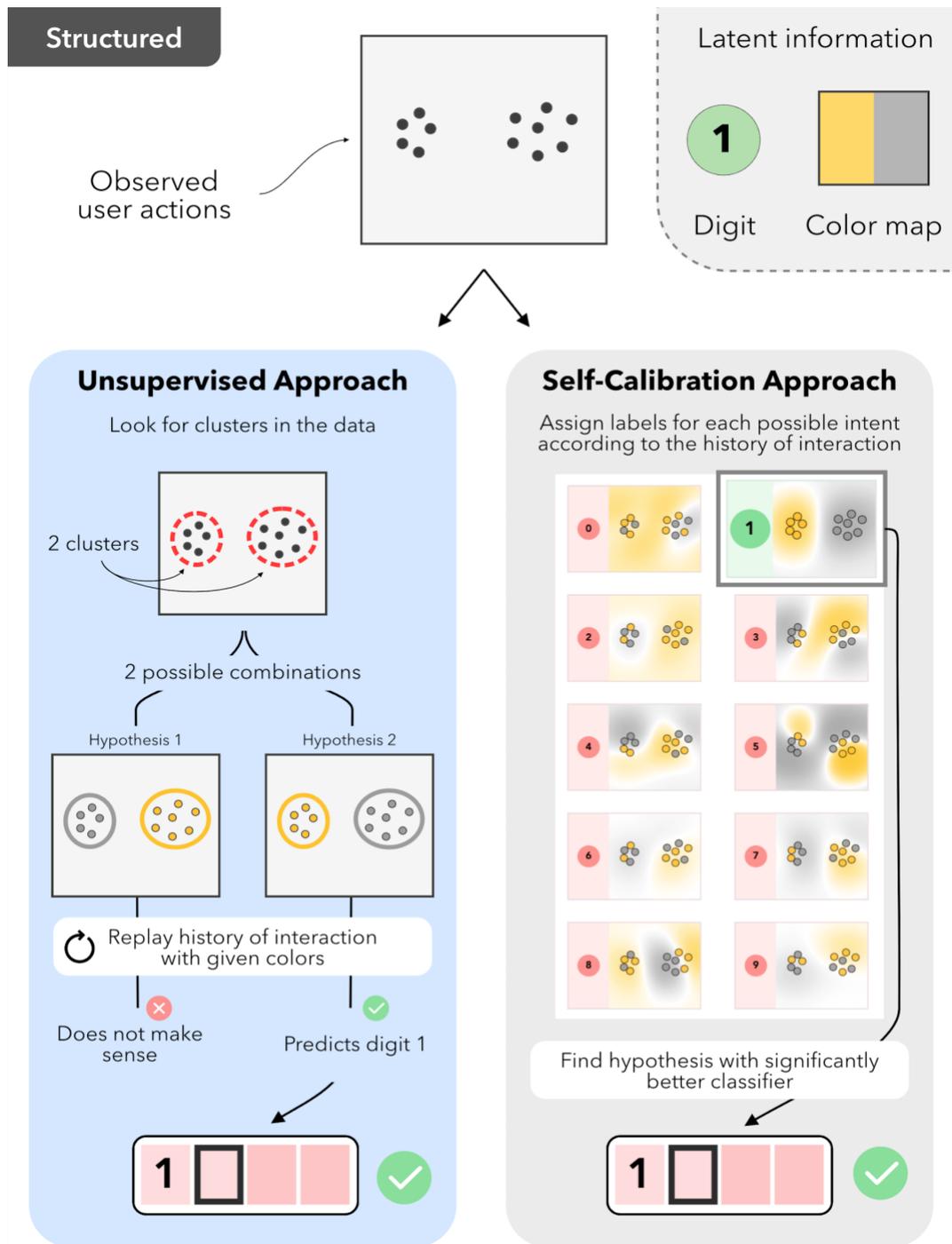

*Figure 20: Comparison of an unsupervised approach (left) to self-calibration versus our method (right). Top-right shows the digit and color map the user has in mind. Top-middle shows the data generated by the user after a few interactions. The unsupervised approach first finds clusters and try different labeling per clusters. The SELF-CAL assign labels according to every possible intent and look for consistency.*

This unsupervised approach is sound and easy to explain because it separates the problem in two logical steps. First identifying potential action-to-meaning mappings and then inferring the user intent given these mappings.

In fact, entire fields of research are dedicated to finding expressive embeddings for all sorts of signals. These embeddings are trained to best split signals by similarity, such that it becomes easy to locate and differentiate them in clusters. Why not then rely entirely on these representations and assume the users' signals will naturally be split into well separated clusters?

Because, despite best efforts, it remains impossible to guarantee that data generated by users will naturally split into clusters, whatever the embedding space. Especially in a self-calibration scenario, where we cannot know in advance the distribution of the signals the user will choose to use. It is thus impossible to engineer in advance a feature space that will ensure that the data will form well separated clusters.

The advantage of the SELF-CAL method is that it does not need to assume the data are organized in well separated clusters. It only assumes that the user will preferably use simpler mappings than complex ones.

Two characteristic cases will fail under the UNSUP methodology (Figure 21):

1. **Unstructured data** - When there is no apparent structure in the data, i.e., when there are no visible clusters.

2. **Deceptive data** - When there is a clear structure in the data, but this structure is deceptive and does not map with the underlying class distribution.

It is important to realize that, in both cases, if we had access to the underlying labels, we would be able to train a classifier capable of differentiating between classes with perfect accuracy.

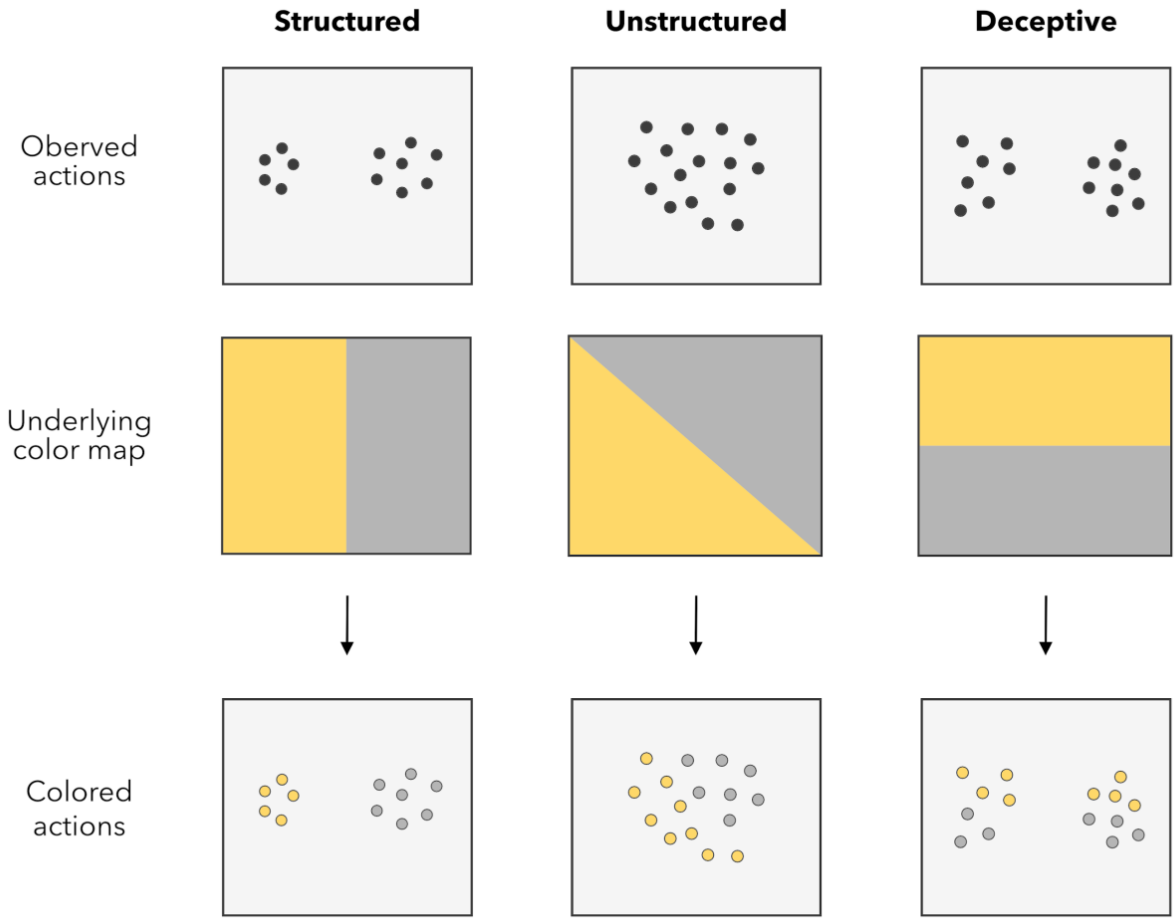

*Figure 21: Illustration of different data distributions to consider for self-calibration systems. The unstructured (middle) and deceptive (right) cases will likely fool the UNSUP approach.*

**Unstructured data**

With this in mind, we first analyze the unstructured case in detail and Figure 22 will illustrates how both UNSUP and SELF-CAL would perform on an example unstructured dataset. The user data is generated from a single cluster in the middle of the feature space. But the color mapping is split diagonally, with the upper-right area associated with grey and the bottom-left area with yellow (top-right of Figure 22).

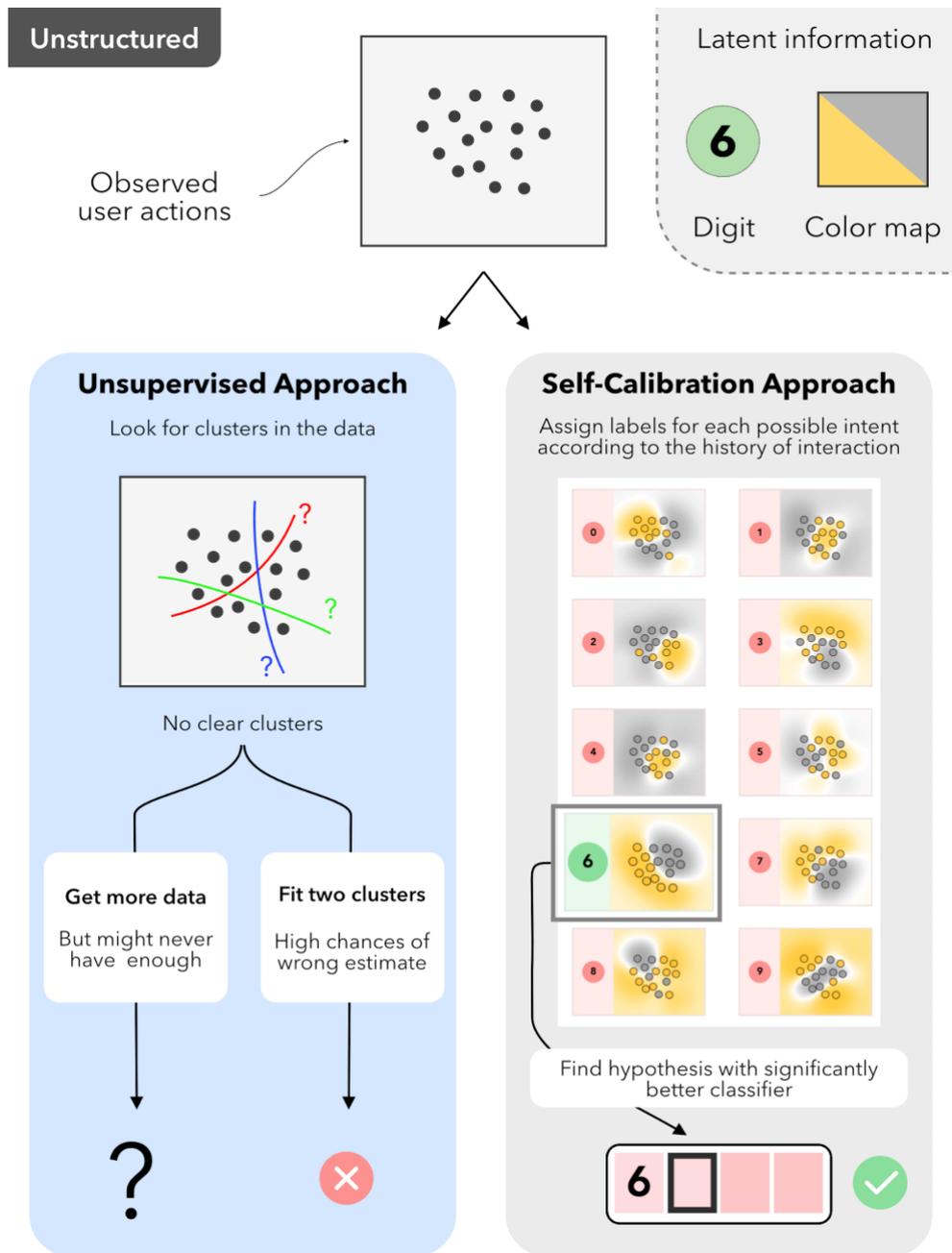

*Figure 22: Illustration of how UNSUP and SELF-CAL would deal with unstructured data.*

Using UNSUP (Figure 22 - left), one would be hard pushed to identify two clusters from the unlabeled data generated by that user. Running an unsupervised clustering algorithm on these data would most likely lead to a wrong clustering. And waiting for more data would not help either because the data are generated from a single Gaussian distribution, not two.

Using SELF-CAL (Figure 22 - right), the problem can be solved because we leverage constraints coming from the task. We know that the user is typing one out of ten digits. Thus, for the same data, we can generate 10 hypothetical labeling, one for each digit. We then simply find the labelling system that is significantly more consistent with the data, as described earlier in this section.

**Deceptive data**

The deceptive case is illustrated in Figure 23. The user data is generated from two horizontally separated clusters, one on the right and one on the left of the feature space. But the color mapping is split vertically, with the upper area associated with yellow and the bottom area with grey (top-right of Figure 23).

Using the UNSUP (Figure 23 - left), two clusters can easily be identified. However, replaying the history of a user's action assuming one cluster is grey and the other is yellow is likely to lead to false prediction or, at best, confusion, and the inability to decide. Waiting for more data would not help either because the data simply are generated using a pattern that does not match with the clustering assumption.

Using SELF-CAL (Figure 23 - right), this problem can be solved because we have access to a limited set of hypothetical labels. The top/bottom split associated with digit 5 is a more consistent mapping than any of the other labeling systems. Note that none of the alternative labeling is considering a strict left/right split, the closest is for digit 9 but one point on each side is of the opposite color which, in that specific case, was enough to discard digit 9 compared to digit 5.

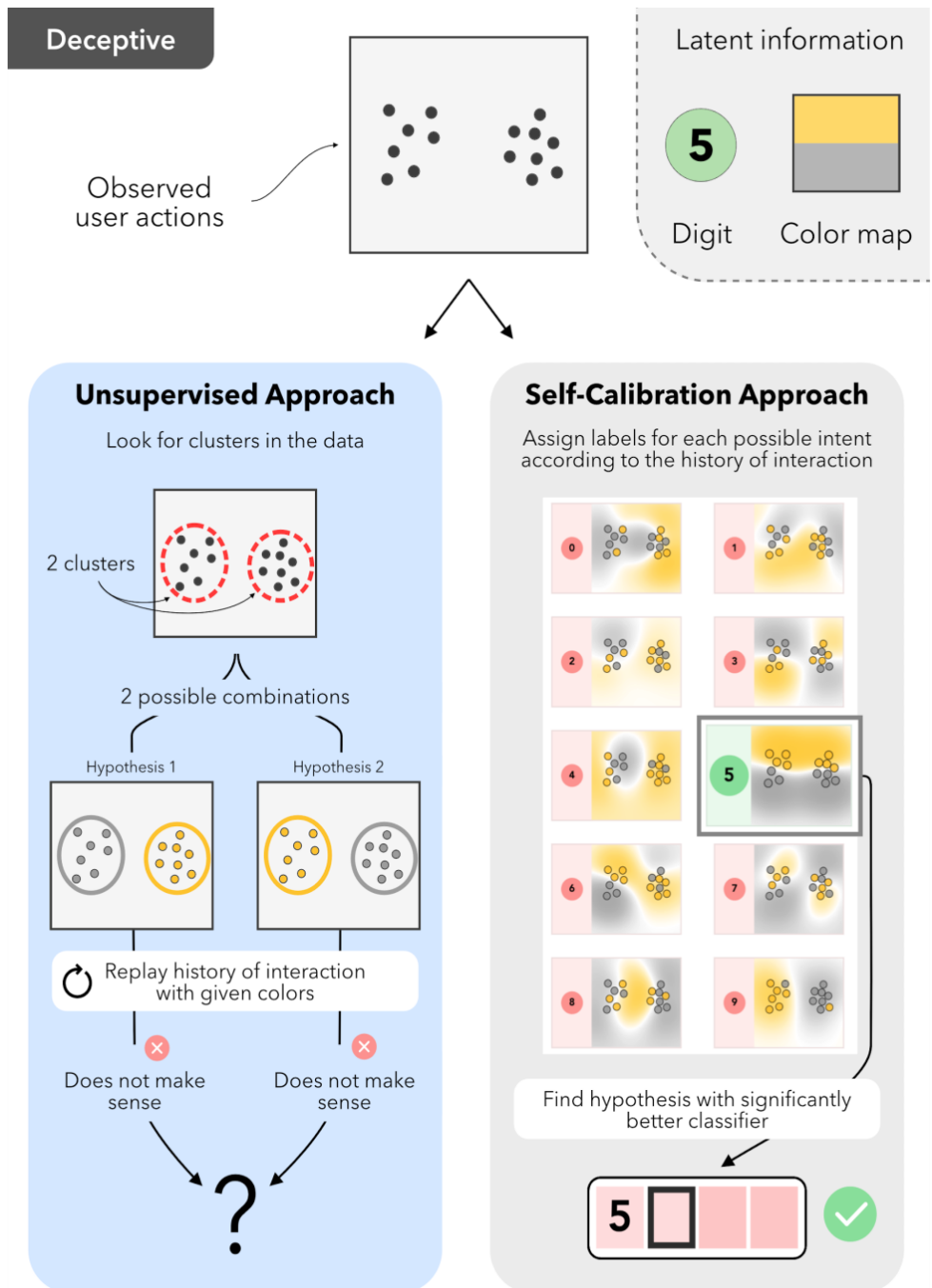

*Figure 23: Illustration of how UNSUP and SELF-CAL would deal with deceptive data.*

SELF-CAL works in unstructured and deceptive cases above because it does not assume data should form clusters or have "visible" structure in the feature space. We only assume that, if we had access to the ground truth label, it would be possible to train a classifier differentiating each class.

This is an important difference, and it is what fundamentally separates the field of unsupervised and supervised learning. In unsupervised learning, we ask: "Can we find clusters in the data?" In supervised learning, we ask: "Can we separate the yellow and the grey points? ".

At a first glance, self-calibration problems seem to belong to the unsupervised learning category because there is no direct way to assign ground-truth labels. Indeed, if we do not know the meaning of an individual user action, then we cannot assign a yellow or grey color to each point.

But, because of constraints coming from the PIN entering task, we can have access to hypothetical labels. It allows us to transform the problem into a comparative study between a finite number of supervised learning problems. We only need to pick the labelling system that is more consistent with the data we collected, and do not have to rely on clustering analysis to do so.

We are using very loose terms here due to the nature of this article. More precise definition, notation, test cases, theoretical and experimental evidence are needed to decipher this point better within an actionable theoretical framework. Attempts have been made and we link to the corresponding literature in section 6.1. For practitioners interested in this direction of research, a lot of work remains to be done with plenty of room for innovation.

In the meantime, we strongly encourage you to challenge the interface by yourself. It is the best way to forge an intuitive understanding of what we described in this section. For example, try generating data whose spatial configuration does not straightforwardly match with the underlying color mapping you arbitrarily choose.

## 4.2. We do not seek to classify user's actions into their meanings

Here, we aim to convince you that an action-to-meaning classifier is only a byproduct of our approach, not its goal. In fact, compared to a traditional human-machine interaction pipeline, we never classify individual user's actions into their meanings. We rather take a global approach by looking for the most consistent hypothetical labelling system.

In other words, instead of asking: "What does a user mean when they perform such or such action?" and then inferring the user intent from those meanings. We rather ask: "Which set of hypothetical labels fit best the data we received from the user?" and we directly identify the user intent that way. Which in turn can inform us of the meanings of each user action. This is a significant shift in approach from the traditional human-machine interaction paradigm.

Why does this framing matter? Because it explicit that, using SELF-CAL, we never commit to a definite action-to-meaning mapping. SELF-CAL allows our model of the user to change in time.

To explain this, we need to differentiate two stages in the learning process:

- Stage 1 - before we identify the first digit
- Stage 2 - after we identify the first digit

**Stage 1**

In stage 1, the difference with a calibration first approach is obvious. With the CALIB approach, we need a classifier to be able to interpret the user's actions. To train this classifier, a calibration step is performed first, where labeled data are collected using a known protocol and a classifier is trained on these data. Then, the classifier is "frozen" and used to translate actions from users into their meanings.

This classifier is thus unique, pre-trained, and frozen in time. It is supposed to be an accurate action-to-meaning mapping of the user. However, the user might still generate signals out of the range of the data observed in the training set, potentially leading to false prediction. More alarming, because there's no way to know about this mistake, the classifier will never be updated with this new data and the problem will reoccur in the future.

With SELF-CAL, we start without any information, no classifier, no data. We compare the consistency of hypothetical classifiers trained only on the data we received and decide only when we are confident about the user intent.

This can happen at any time, and we might not have acquired enough data to cover all possible user signals, we collected just enough evidence to be confident of the intended digit. This means that the color map we have is unlikely to be accurate and would inevitably lead to wrong predictions if we were to "freeze" it and use it as a calibrated classifier.

Take the example in Figure 24. We simulate a user that defined three areas on the color map - Left/Middle/Right and is typing the code 2020. For the first digit, a 2, the user places points in the left area for yellow, and the middle for grey, but never uses the right area.

Once the machine identifies the first digit, a 2, we could be tempted to train a classifier on the associated labels. And because the user never used the right area, the best guess from the classifier would be to consider it as a grey area. Simply because it is on the side of the grey points.

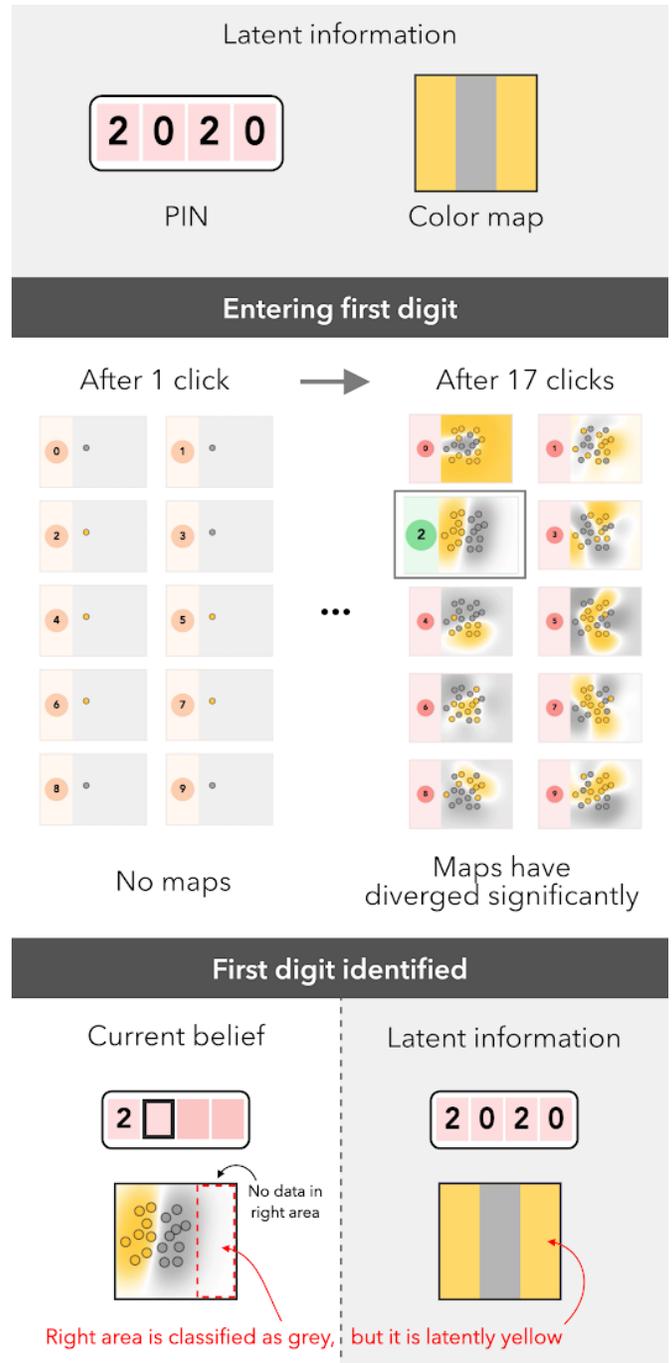

Figure 24: Issues arising when we freeze a classifier before we have received representative signals from the user. Using CALIB approach, we would us the color map on the bottom-left side as the ground-truth classifier and mistakenly predict points on the right of the map to be grey. We will see how SELF-CAL avoid this issue.

**Stage 2**

A key component of SELF-CAL is that, once a digit is identified, we do not freeze the associated classifier to classify subsequent user's signals. In facts, we never use any classifier to directly classify user signals. Rather, we propagate the labels associated with the "winning" hypothesis to all other hypotheses (because we are now confident that the labels associated with the user intended digit are the ground truth for the signals received so far) but every new signal will continue to be handled using the SELF_CAL method.

Figure 25 illustrated the propagation progress. On the left are the color maps for each possible digits right before digit 2 is identified as the user intent. Immediately after, the labeling associated to digit 2 is used for all other maps. As a result all maps looks the same at this precise point in time.

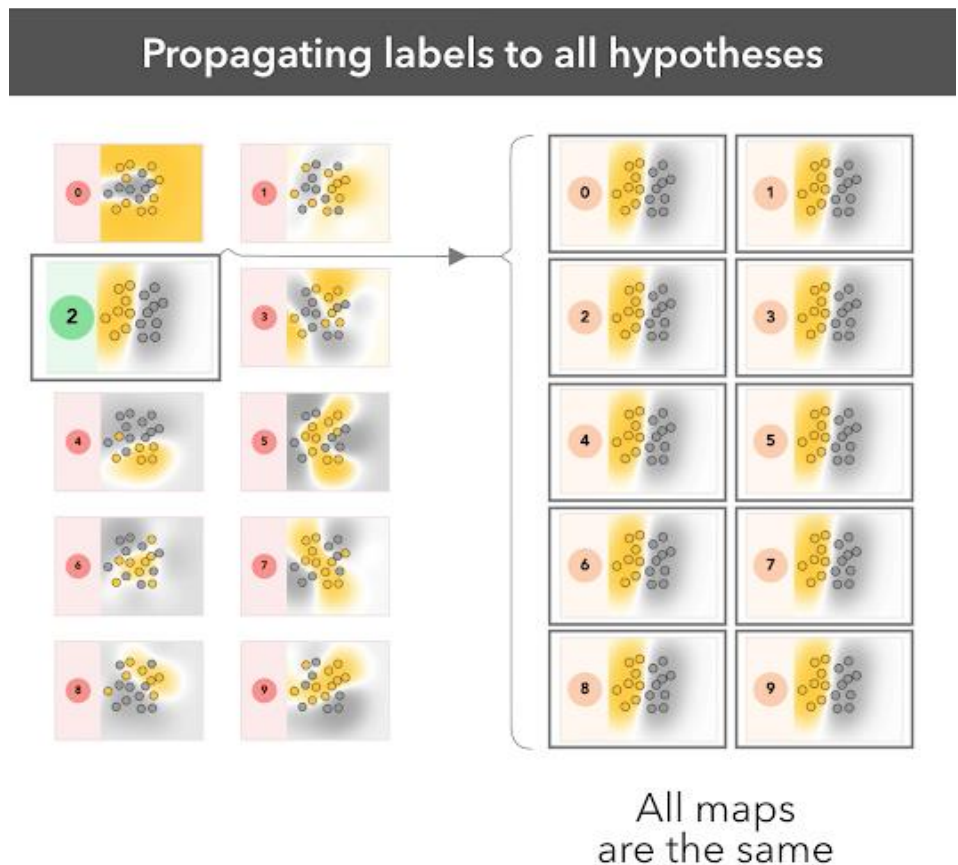

*Figure 25: Label propagation after a digit has been identified. At this point in time, all color-to-meaning mappings look the same. They will start to diverge again with the next incomign signal.*

But, as exemplified in Figure 24, we should not rely on the fact that, now, all classifiers are the same because they might still be terribly wrong, especially on data coming out of observed distribution. Instead, SELF-CAL rather consider previous data as a prior, as a valuable source of information but not as representing the ground. In practice, we simply continue the hypothetical labelling procedure and assign subsequent user's actions with different labels according to each hypothesis.

This process will drive the hypothesis classifiers away from each other's again and a new decision will be made when one of the classifiers is significantly more consistent and more likely to explain the user behavior. Because we continue assigning labels according to the hypothetical intents, rather than referring to a frozen classifier, the action-to-meaning mapping is continuously updated. This is particularly advantageous in locations of the feature space where no data was collected before.

To illustrate this point, we continue our example of Figure 26. Once the first digit is found, the user decides to enter its second digit, a 0, by continuing using the middle area for grey, but by now using the right area for yellow, and never using again the left area.

If we had frozen the best classifier learned for the first digit, all clicks in the right area would have been predicted to be grey, and the digit predicted would have likely been wrong.

With the label propagation trick, the machine was not lead ashtray, did not over generalize, identified the correct digit, and correctly learnt that the right area of the map was used to mean yellow by the user.

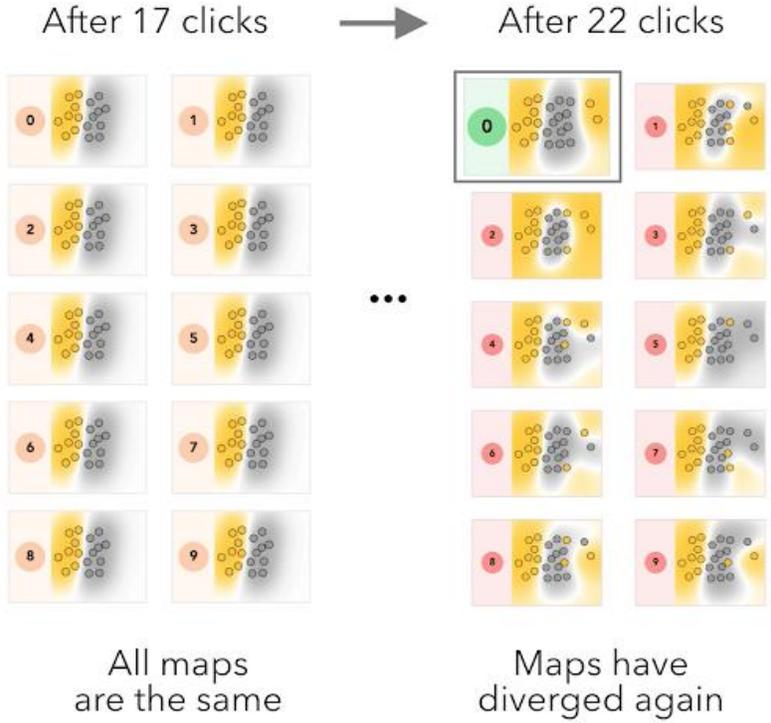
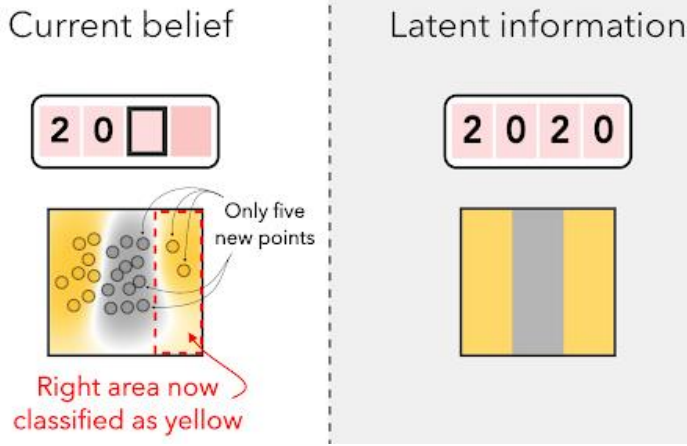

*Figure 26: Continuing SELF-CAL procedure after a first digit is identified continually updates the possible action-to-meaning mappings and maps for all digits diverge again. After a few more clicks, the digit 0 clearly has a more consistent mapping and is thus selected as the one the user has in mind (top-right). Notice how, with only five news points, we identified the new digit correctly and update our color-to-meaning map meaningfully.*

The same process of label propagation is then repeated for subsequent digits. Interestingly, as more digits are entered, the identification of the next digit becomes faster because the amount of prior information shared between each hypothesis increases each time a digit is identified, and labels are propagated.

To continue our example, after the third digit is identified, 28 points share the same labels for all hypotheses (Figure 27). For the last digit, if the user clicks in the center of the 2D map, five hypotheses will assign yellow to this new point and the remaining five will assign the color grey. But because this point is surrounded by grey points for all hypotheses, half of the hypothesis (the one assigning it to yellow) will almost immediately be discarded.

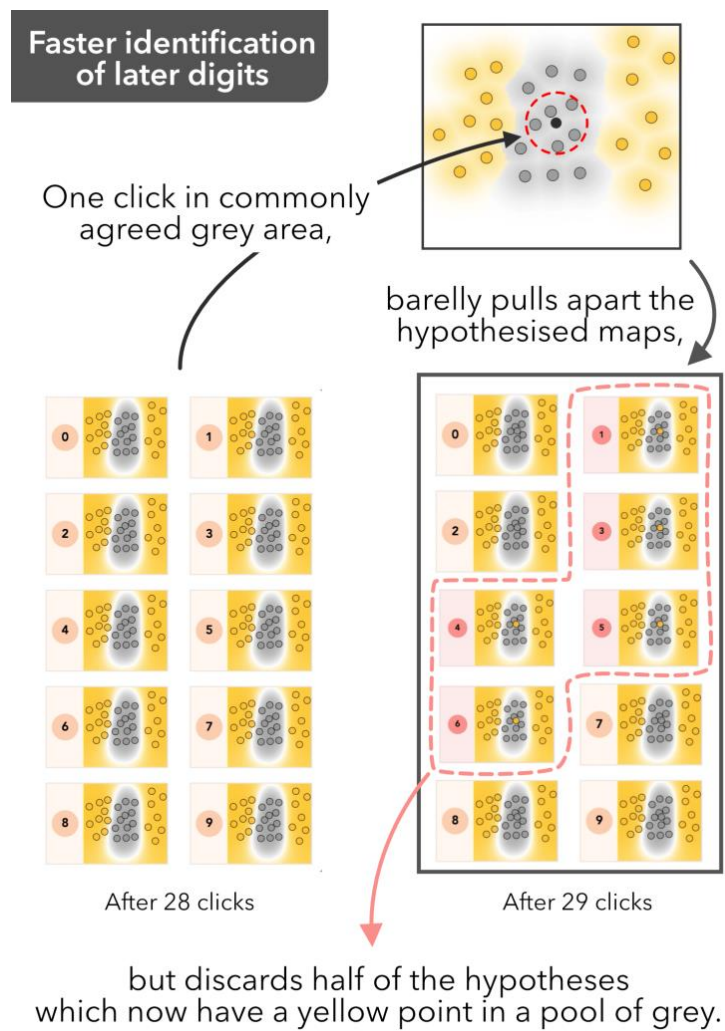

*Figure 27: Compounding effect of interaction with SELF-CAL. The previous data acting as a prior, the more previous interaction the user had with the machine, the more meaningful are signals received by the user (if those signals are received in an area where we received signals before).*

SELF-CAL is thus a gradual process that starts from knowing nothing about the action-to-meaning mapping of the user to having a very good model of it. We could say that stage 1 starts with no prior information about the user data[5]. While in stage 2, concrete prior information is available in the form of the ground-truth labels of some data. The more digits we identify, the stronger the prior. A larger and larger proportion of signal-label pairs are shared between all hypotheses, making it easier to detect inconsistencies. One yellow point landing in the middle of a pool of 20 grey points shared by each hypothesis becomes a strong sign of a breach in consistency.

Following this logic to infinity, the SELF-CAL procedure is progressively converging to the CALIB methodology. Once all hypotheses share hundreds of points with the same labels, a new point, whatever its label, is unlikely to lead to significant changes between the hypothetical classifiers. At this point, all hypothesis acts as one unique classifier, exactly like the CALIB method.

It is worth noting again that we reached that stage without ever explicitly predicting meanings from actions, we rather compared alternative classifiers. This transition from pure self-calibration to fully calibrated is nicely embedded within the SELF-CAL framework. It illustrates that, seen from a new angle, the reasoning behind the CALIB problem is a limit case on the SELF-CAL spectrum where all hypotheses share the same infinite prior and agree on the color map used by the user.

I hope that I have convinced you that SELF-CAL does not seek to train an action-to-meaning classifier at all. By not trying, we are more flexible and more robust to edge cases or novel signals arising from the user. Naturally, if one really wants to train a unique action-to-meaning classifier, it can be done at any time. SELF-CAL gives access to the ground-truth labels after each digit is identified, thus we can train a classifier using this shared prior. But this is only a side effect of the method, not its intent.

---

[5] Although it is good to remember that a lot of assumptions are made all along, such as the user following the protocol, the data can be classified with the classification algorithm selected (here SVM with RBF kernel which embeds its own assumptions about the structure of the data), and the assumption of consistency stating that the user prefers simple maps to complex ones.

## 4.3. Our stopping criteria is both data and task dependent

We call "task" the problem to be solved in interaction with the user, here entering digits of a PIN. We call "data" the collection of actions communicated by the user via the interface, for example the history of buttons clicked, or the points placed on the 2D maps.

Two conditions need to be met to decide which digit the user intends to type:

1. **We need enough information to solve the task.** We cannot decide until all digits have pairwise been of a different color at least once. For example, if the user is typing a 0, but the digits 0 and 1 are always of the same color, it will be impossible to decide whether the user is typing a 0 or a 1 because the actions of the user will always be consistent with both 0 and 1. The method we use to select the color of the digit (more on this in section 6) is designed to make sure this does not happen but it is important to remember it is a component of the problem. Therefore, if we increase the number of intents, for example typing letters instead of digits, more clicks will be required to identify the user intent. Going further, tasks could be multi-step processes, such as playing a video game or navigating a maze. In such cases, the agent needs to reach very specific states to be able to eliminate some hypothesis, which might require long sequences of action. In interactive learning scenarios like the one we present here, there is always a lower bound in the number of interactions required to solve the task.

2. **We need enough information to understand the data.** All hypothetic classifiers remain equally valid until we have collected enough clicks to identify some structure in the data. A good way to understand this is to refer to the interface with buttons from section 2. Knowing the color of the buttons, it takes 3 or 4 clicks to identify a digit. But, if we do not know the color of the buttons, we need at least 2 clicks on one button to start finding breaches in consistency and eliminate some hypotheses. If a user clicks only once on each button, we will collect 9 clicks in total, yet no digits could be eliminated. All hypotheses would remain valid, each having a different but consistent model of the user action-to-meaning mapping. There is always a lower bound in the quantity of data required to identify some structure in the data.

The SELF-CAL algorithm is solving the task at the same time as it is understanding the data. If the task is hard to solve or the data is hard to understand, SELF-CAL will automatically account for it and not make any decision before a confidence threshold is reached. Thus, because both task and data constraints are embedded into the same algorithms, SELF-CAL offers a smooth transition to a problem that is first limited by the understanding of the data (Stage 1), and then, once enough prior information on the data is available, limited by the task (Stage 2).

Data constraints dominating in Stage 1 are best explained using the interface with buttons. With calculated strategies, it is possible to never let the machine know what our digit is. To do so, you need to focus on a few digits and carefully click on buttons such that the action-to-meaning mappings associated with all these digits remain consistent over time. Let's try it on https://openvault.jgrizou.com/#/ui/tuto_3x3.json using the pattern from Figure 28.

Focus on two digits and four buttons. For example, only look at digit 0 and 1 and at the four most top-left buttons. Associate a unique button to each possible color combination of the digit pair. There are only four possible combinations, as exemplified in Figure 28 (left). Using the interface using this pattern of action ensures that, for all possible coloring applied to digit 0 and 1, you always press the same buttons and remain consistent.

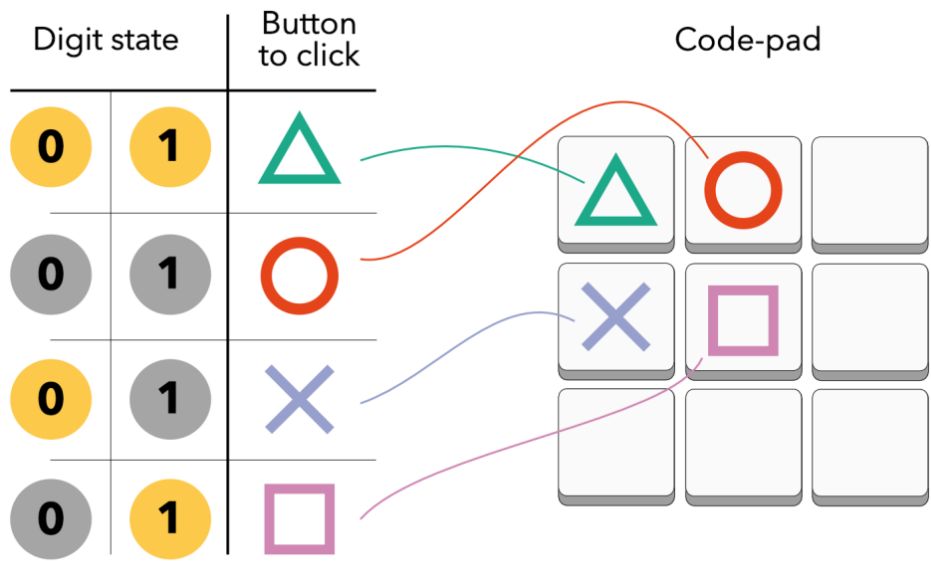

*Figure 28: Pattern of button clicks testing the limit of SELF-CAL decision making process. If you use the buttons in that way, our interface will never be able to know which of the digit 0 or 1 you intend to enter. You would indeed be acting consistently with respect to two different digits.*

Try using the interface while following this logic. Figure 29 shows the button-to-color maps associated to digit 0, 1, 2, 3 after more than 36 clicks following this process. It is also demonstrated in this video https://youtu.be/5HpDeInQc_w. It is impossible for the algorithm to decide between the digit 0 and 1, both usages of the button are equally plausible. However, it is easy to see that digits 2 and 3 are not the one the user is trying to enter due to irregularities in the meaning associated to each button.

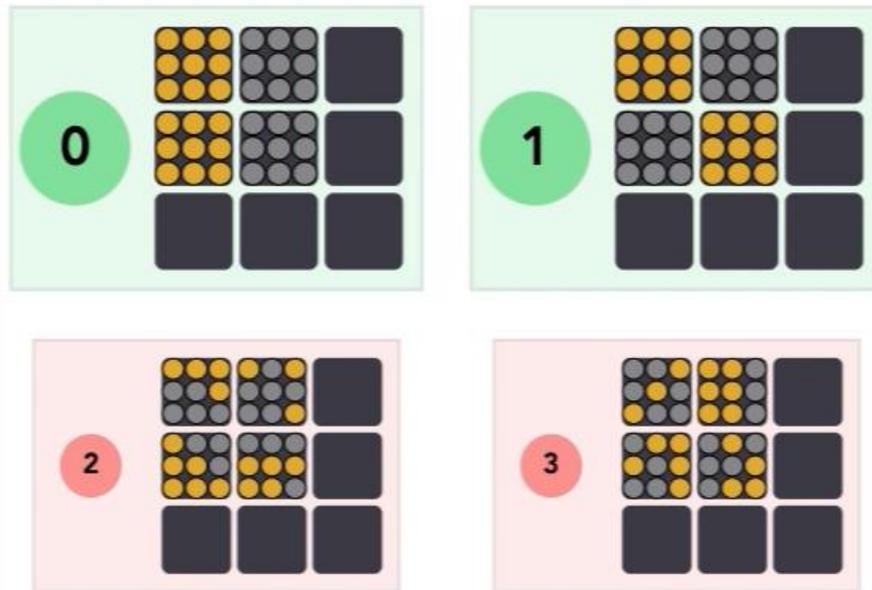

*Figure 29: Result of following the click pattern from Figure 28 on the IFTT-PIN button-based interface. Despite clicking 36 times, SELF-CAL could not decide whether the user was typing a 0 or a 1. Indeed, both options remain as consistent after 36 clicks. See https://youtu.be/5HpDeInQc_w for a demo.*

This result is somewhat surprising, you can click as long as you want, and the machine will never be able to identify what digit you want to type. Now think again, do you even know what digit you are trying to type? Maybe not. You are rather purposefully trying to trick the machine. Nonetheless, it is theoretically possible for a user to try to type a 0 and, by luck, use the same four buttons in the same way as you did. However, it is very unlikely for this to happen without a conscious effort to trick the interface.

The same logic applies on the touch version of IFTT-PIN. Instead of selecting four buttons, you can split the screen in four areas and associate one area for each digit state, as shown on Figure 30.

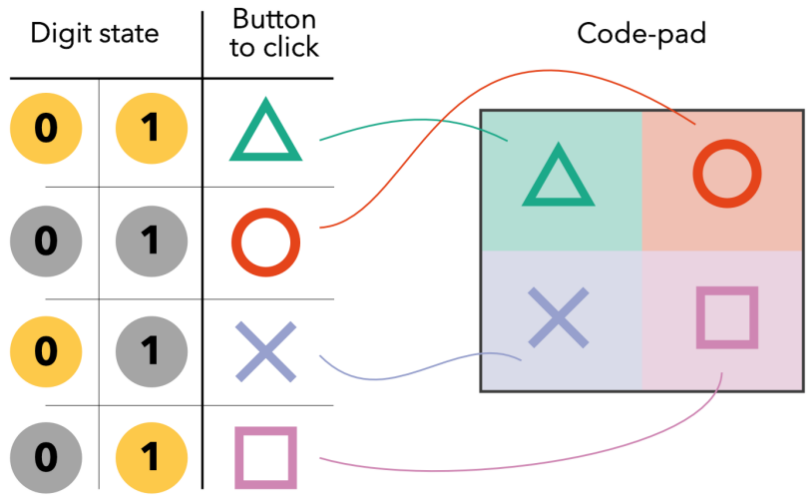

*Figure 30: Pattern of button clicks testing the limit of SELF-CAL decision making process. If you use the map in that way, our interface will never be able to know which of the digit 0 or 1 you intend to enter. You would indeed be acting consistently with respect to two different digits.*

Let's try it on https://openvault.jgrizou.com/#/ui/tuto_touch.json using the pattern from Figure 30.

Figure 31 shows the classifiers associated to digit 0, 1, 2, 3 after more than 50 clicks following this pattern. It is also demonstrated at https://youtu.be/Gf48wNd1W4k. It is impossible for the algorithm to decide between the digit 0 and 1, both maps are equally consistent. However, it is easy to see that digits 2 and 3 are not the digit the user is trying to enter.

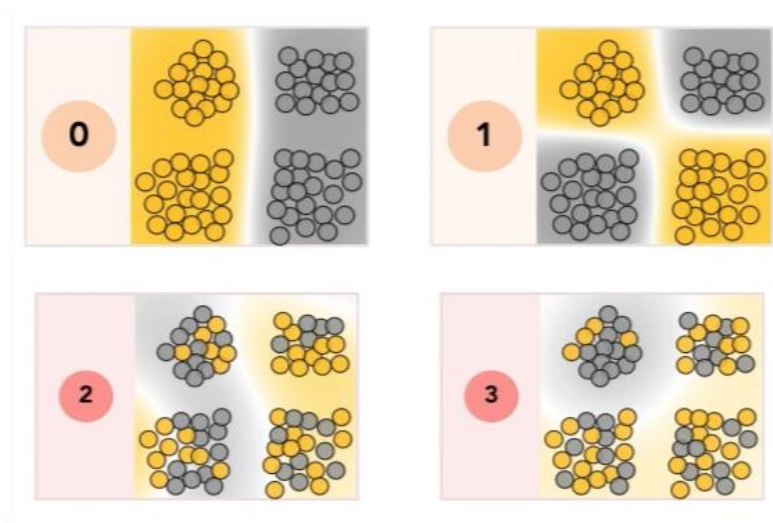

*Figure 31: Result of following the click pattern from Figure 30 on the IFTT-PIN touch-based interface. Despite clicking 36 times, SELF-CAL could not decide whether the user was typing a 0 or a 1. Indeed, both options remain as consistent after 36 clicks. See https://youtu.be/Gf48wNd1W4k for a demo.*

The above demonstration emphasizes how Stage 1 is limited by the understanding of the data. In Stage 2, once one or more digits are identified, the process becomes mostly limited by the task. Indeed, even if we had access to a perfect action-to-meaning mapping, the algorithm would still need 3 or 4 clicks to pull apart each hypothesis.

The SELF-CAL procedure nicely adapts throughout Stage 1 and Stage 2 of the interaction as the weight of the prior information progressively increases. Unlike the CALIB methodology, our method does not need to rely on an explicit calibration phase to first understand the data and, only once the data are understood, focus on the task assuming the action-to-meaning mapping is correct.

Indeed, a problem with the CALIB approach is to know when to stop the calibration procedure. How can we be sure that the trained classifier represents the future user behavior well? Maybe we should ask the user to repeat the task a couple more times to get a bit more data. Maybe we already have too much data and wasted everybody's time. SELF-CAL removes this problem because it merges both the task and the data problems into one.

SELF-CAL will stop when it has just enough information to be confident of the user intent. No more, no less. Hence, contrary to a CALIB procedure, we do not even need to have a good estimation of the action-to-meaning mapping used by a user to identify their intent. We only need to have enough information to pull hypotheses apart. In our demo, the first digit is often identified before enough data is available to build a thorough color map. As we saw in subsection 4.2 above, the map can actually be wrong and corrected later thanks to the hypothetical labelling procedure.

---

Section 3 and 4 hopefully make a compelling case of why SELF-CAL scales well to user actions represented in continuous spaces. If we can solve the self-calibration problem with 2D points on a map, it should also work with a wide range of modalities in higher dimensional spaces.

In the next section, we demonstrate the use of drawings and spoken commands to enter a PIN in our interface. Once again, you will get to decide which drawing or spoken word is associated with yellow or grey. You will invent a simplified sign and spoken language, which the machine will learn without prior knowledge, training set or calibration phase.

# 5. Draw and speak

The goal of this section is to demonstrate the use of higher-level modalities within the self-calibration paradigm. We will use hand-sketches and spoken commands to interact with IFTT-PIN.

As you can intuit, hand-sketches and spoken commands are harder to work with than points on a 2D map. A sketch could be represented as the list of all the positions a pen goes through. A spoken word could be represented as the list of all the amplitudes recorded by your microphone in time sampled at several kHz.

In other words, instead of working with 2 dimensional vectors encoding a [x, y] position on a map, we now deal with vectors of N dimensions with N likely to be very large, which makes it harder to find patterns in the data. Luckily, the scientific community spends a lot of time trying to come up with compact representations for various types of signals.

[Feature extraction](#) is the process of describing a phenomenon of interest using a limited number of characteristic features while conserving relevant information. These compressed representations are helpful to visualize and interpret the data. For example, in medicine, the height and weight of an individual can be used to predict risk of cardiovascular diseases, see the [body mass index](#) for example. Summarizing a human being by its height and weight is an example of feature extraction.

In machine learning, feature extraction facilitates learning and generalizing. Instead of working with data in large dimensional spaces, we extract a few key features from the data and use these features as representing the whole for our analysis. The challenge is to extract the right features that conserve key information about the original data such that the desired task can be solved to satisfaction.

A complementary method is [dimensionality reduction](#) whose goal is to project data from a N-dimensional space into a smaller D-dimensional space while conserving the relevant information and relationship between the data. Well known dimensionality reduction algorithms include [PCA](#), [t-sne](#) and [UMAP](#).

To scale our approach to drawings and spoken words while conserving our intuitive 2D visualization, we decided to represent drawing and sounds as points on a 2D map using a combination of feature extraction and dimensionality reduction.

While feature extraction can be done on a per sample basis, dimensionality reduction requires a dataset with at least a few samples. We chain both methods and first extract relevant features from each user action (sketches or sounds) and then projects the resulting dataset of signals received from the user in a 2-dimensional space. This level of compression is not usually recommended as a lot of information might be lost in the process. We took that risk in this setup for a few reasons:

- 2D points and color maps are easy to visualize and we want you to be able to follow the process using our interactive dashboard.
- We start from scratch and have only two classes so we do not have to learn a representation that can cluster all possible drawings or speech but only the one generated by users, making the task easier.
- We do not rely directly on a classifier trained on the projection but use the SELF-CAL procedure which, as shown in section 4, is robust to deceiving dataset distributions.
- Users do not have infinite patience and we need to identify a digit quickly. The higher the dimensionality, the more data is required for SELF-CAL to identify your digit. By working on 2-dimensional data we ensure a decision is reached quicker.
- Making a mistake has no real bad consequences. The worst that can happen is a false digit showing up on the screen, which we hope the reader will understand.

Next, we will provide an overview of the pipeline used to project sketches and sounds in a 2D space, but we will not go into details to maintain the focus on the self-calibrating concepts and its implications. The goal here is for you to experience the interface using sketches and sounds.

## 5.1. Hand-sketches

We represent sketches using simple features such as the start and end [x, y] coordinates, the distance between the start and end location, the length of the drawing path, and the percentage of the drawing falling into each cell of a 3x3 grid - forming a 17D feature vector for each sketch. To be robust to scale and location, a normalizing step is applied before features are extracted[6].

---

[6] Our representation is robust to differences in scale and position, but not in rotation. For example, using squares of any orientation for yellow and triangles of any orientation for grey might not work. It could be a good experiment to challenge the system though.

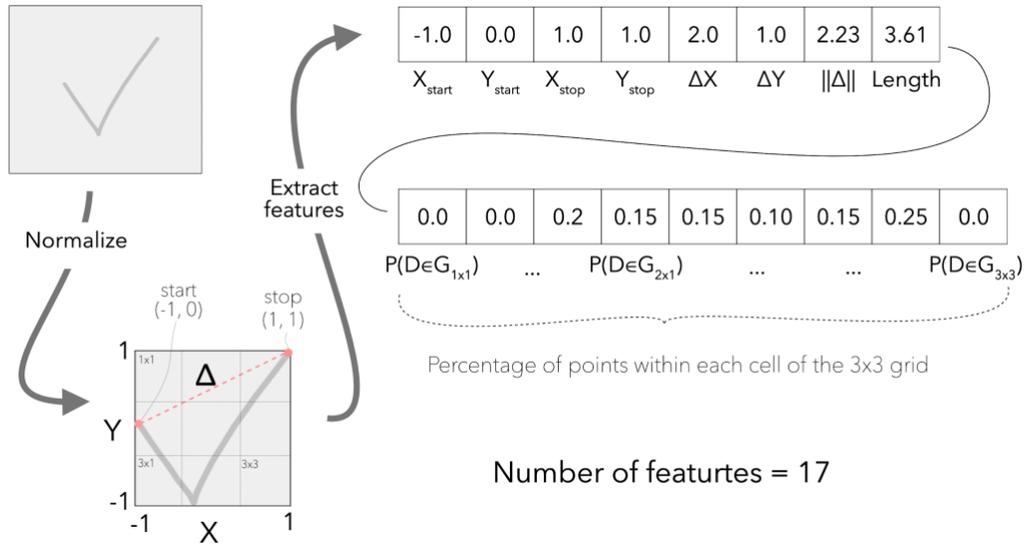

*Figure 32: Illustration of the feature extraction process of hand-sketches. The sketch is first scale to fit into a [-1, 1] grid. Various features are used to represent the sketch such as start position, end position, delta position on the X and Y axis, as well as the percentage of the drawing falling into each cell of the 3x3 grid.*

All sketches received from a user are then collated in one dataset and the UMAP algorithm is used to project the data from the 17D feature space to a 2D space.

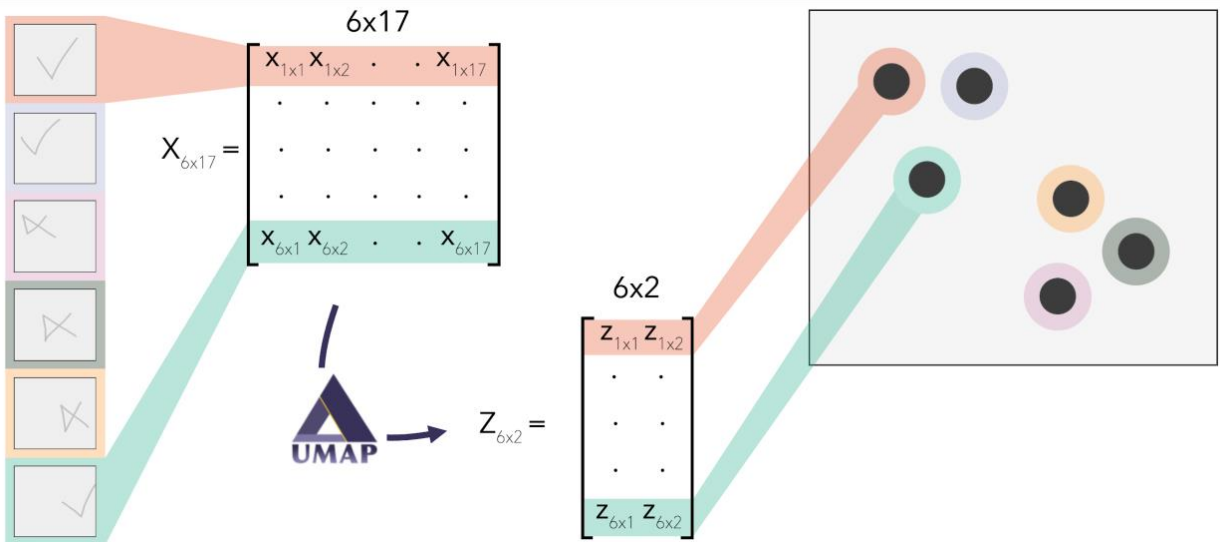

*Figure 33: Illustration of the dimensionality reduction step. In the example, 6 sketches are first represented using the features extraction into 6 rows of 17D vectors. The UMAP algorithm is then applied to reduce dimensionality to 2 allowing us to plot each drawing as a point on a map, similar to what we saw in section 3.*

As we are in a self-calibration scenario, you decide which drawings to associate to yellow and to grey. It is arbitrary and up to you. For example, a triangle could mean yellow, and a circle mean grey. The drawing of a house could mean yellow, and a carrot mean grey. You decide. Maybe a safe place to start is to draw the letter 'Y' for yellow and the letter 'G' for grey, simply because it is easy to remember. And don't panic, your drawings do not have to be accurate or pretty, rough sketches are perfectly fine.

A drawing is limited to one stroke of a pen. To start drawing, press the left button of your mouse down on the drawing area. This will drop the pen and you can start drawing with it. Drawing will stop when you release the button of your mouse. The sketch will be automatically sent to the machine as the action associated with the digit state shown on screen.

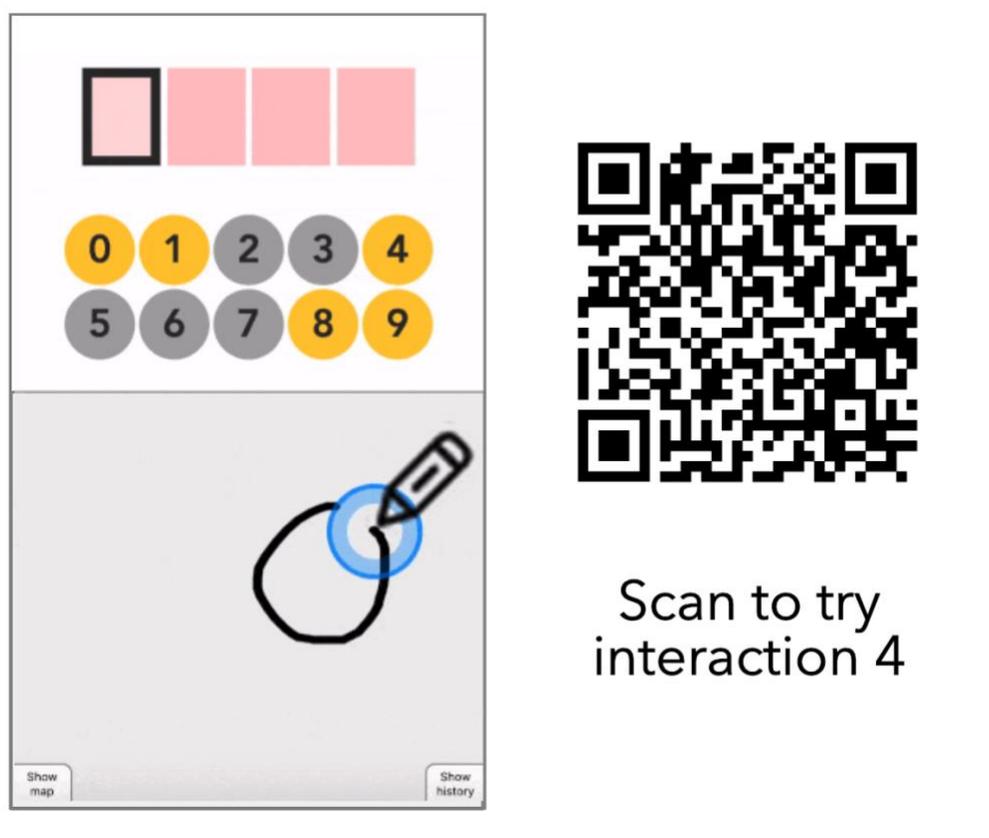

*Interaction 4: The self-calibrating version of IFTT-PIN using hand-sketches as input. The user draws arbitrary sketches on a 2D pad (e.g., circles, crosses, letters, etc.) and decides which sketches means yellow and grey in their mind. Demo available at [https://openvault.jgrizou.com/#/ui/demo_draw.json](https://openvault.jgrizou.com/#/ui/demo_draw.json)*

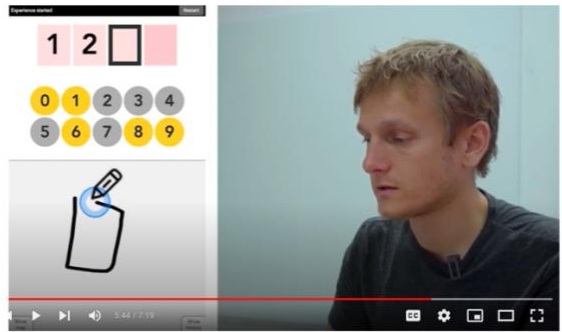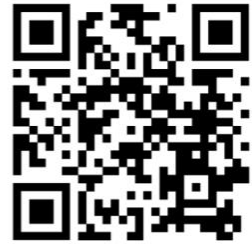

*Video 7: A step-by-step explanatory video of the self-calibrating version of IFTT-PIN using hand sketches as input. Video available at https://youtu.be/5bjk-E8TQLc*

Two new buttons are available at the bottom of the screen. The bottom-right button will show you the history of sketches. This history will be colored in yellow or grey once the machine will have understood what they mean, that is only once a digit has been identified. The bottom-left button will show the projection of your drawing on a 2D space. Note that the UMAP projection is recomputed at each iteration with the new data and will therefore change after each new action from the user. This allows us to refine the projection as more data is available. It does not hurt performances because SELF-CAL estimates the likelihood of each hypothesis from scratch at each iteration.

To visualize the process, you can use the interface with the dashboard as before. The side panel will show you the 2D representation of your label according to each hypothesis. It looks the same as in section 3.

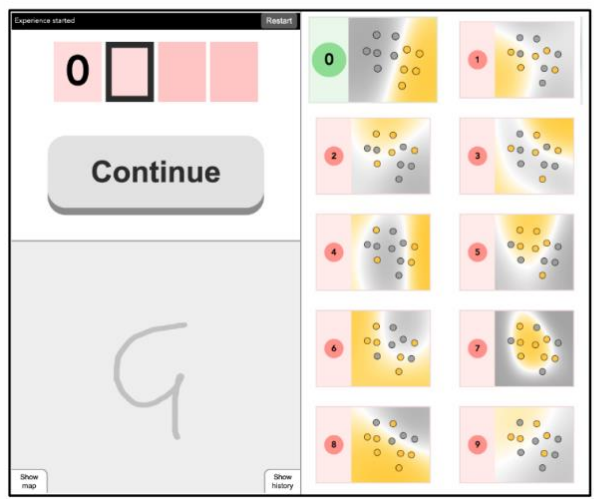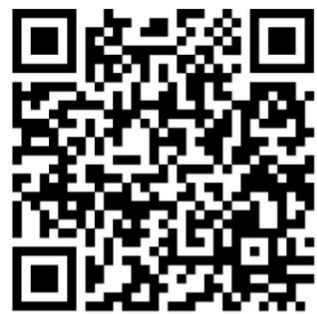

*Explanation 4: The self-calibrating version of IFTT-PIN using hand-sketches as input and with a side explanatory panel. Demo available at https://openvault.jgrizou.com/#/ui/tuto_draw.json*

## 5.2. Spoken words

We applied the same logic to spoken words and represented sounds by pre-trained embeddings. These embeddings were computed from a classification tasks on the [AudioSet](#) dataset (see https://ieeexplore.ieee.org/abstract/document/7952261 and https://arxiv.org/abs/1609.09430) which covers everyday sounds. We decided to use non-speech specific features because users do not have to use spoken words for this task. Self-calibration means they can invent their own language or use sounds generated by objects around them.

The embeddings we use encodes 1 seconds of sounds into a 128-dimensional vector. Because of the high dimensionality of the embedding compare to the size of the datasets we generate using IFTT-PIN, we use data augmentation to help UMAP find structure and clusters with a small amount of data. More specifically, each sound is split into overlapping windows that each create a new entry in our augmented dataset. In other words, we try to create "visible trajectories" in the embedding space by splitting sounds in overlapping chunks. This overlap between samples helps UMAP find manifolds in the embedding space. We then project the entire dataset into a 2D space and average the augmented projections to form the final 2D representation of a sound.

The user can record a sound of at most 3 seconds. The sound is trimmed and repeated to reach a length of exactly 3 seconds. It is then split into 21 windows of 1 second starting every 100ms. Each 1 second's sequence is projected into its embedding of 128 dimensions. See Figure 34.

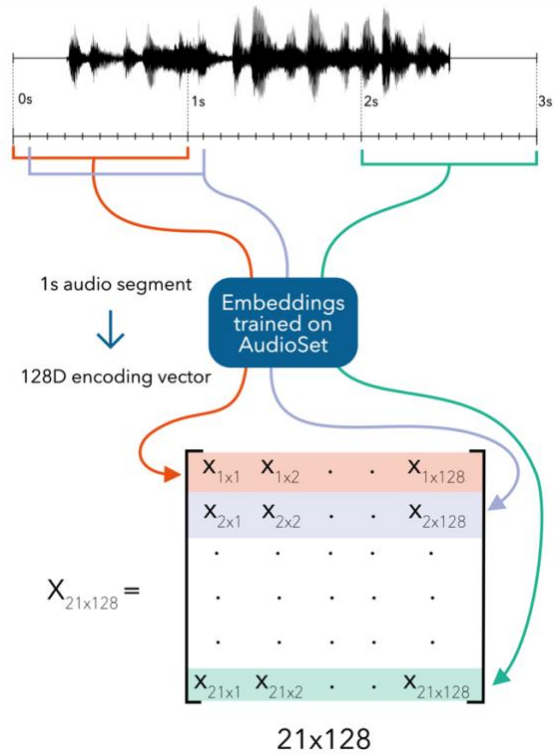

*Figure 34: Illustration of the data augmentation strategy used to represent sounds. A user recording is repeated to fit within a 3 second window. 21 overlapping audio segment are extracted and encoded using the embedding model to create, for each sound, 21 new entries in our dataset, all sharing the same label.*

For N words, we end up with an unlabeled dataset of N*21 entries. We know that a subset of these points is linked together as part of a "sound trajectory" and thus share the same labels. We use the UMAP algorithm to project the data from the 128D into a 2D space. We then average each trajectory projection to come back to a unique projection for each sound which is easier to understand and display on our explanatory interface. It is also the data we use for our SELF-CAL procedure. An example for 6 input sounds is shown on Figure 35.

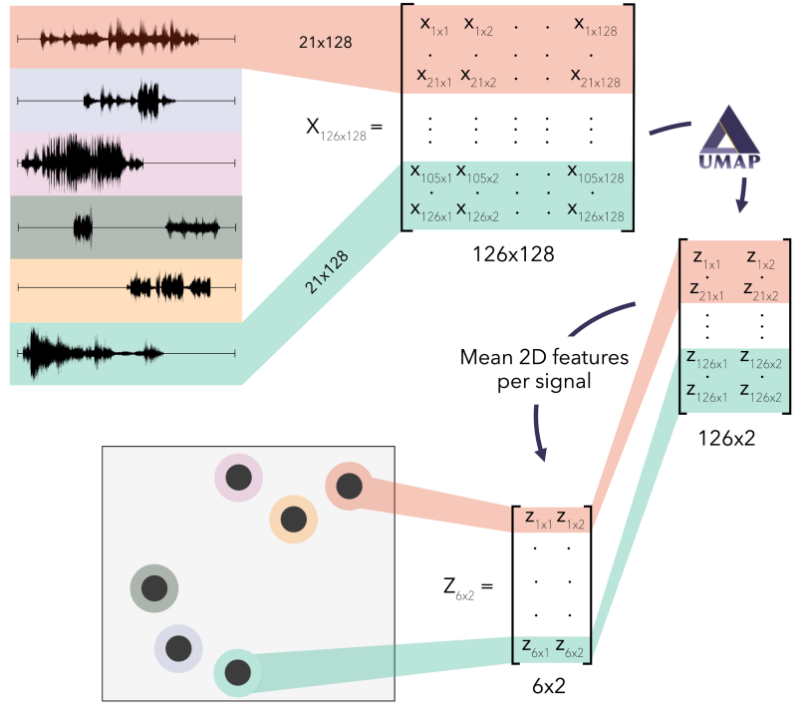

*Figure 35: Illustration of the sounds' representation and dimensionality reduction process. 6 input sounds are embedded into 6x21=126 vectors of dimension 128. The UMAP algorithm is used to project the data to 2 dimensions, from 126x128 to 126x2. By averaging the features from the 21 samples representing one sounds, we end-up with a dataset of 6 entries of 2 dimensions, a very compact representation of our arbitrary sounds that, in most of our testing, contains enough information for SELF-CAL to work.*

This representation is clearly a hack that was needed for a digit to be identified in a few iterations. The use of this generic embedding combined with the trajectory trick to identify manifold allows the user to explore a wide range of sound. It works well for our limited use-case, and we do not recommend relying on similar tricks for problems with real word consequences before thoroughly testing this method. I am sure experts in sound processing could find many ways to improve on this.

It is time to use IFTT-PIN by talking to it by using the interface on Interaction 5. Sounds are limited to 3 seconds in total and recorded as illustrated on Figure 36. To start recording, click on the green button. The recording immediately starts and will stop after 3 seconds. You do not have to make a sound lasting 3 second, we just give you plenty of time to make an arbitrary sound. The sound will be automatically sent to the machine as the action associated with the digit state shown on screen. It will take a few seconds to process, be patient.

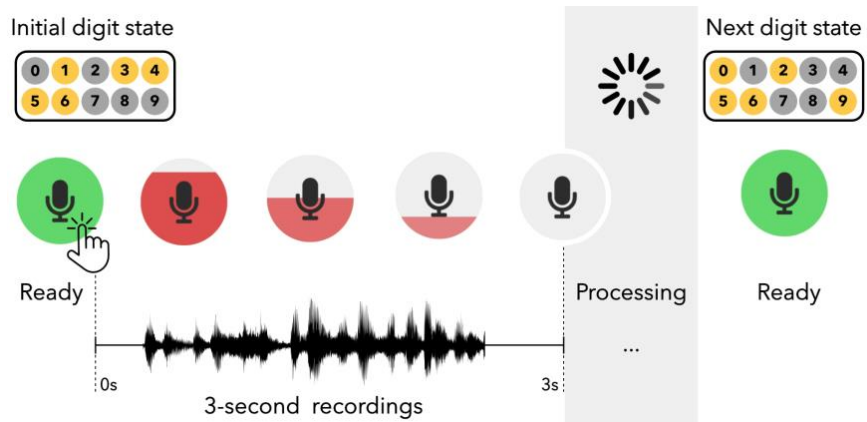

*Figure 36: Illustration of the recording procedure. A sound is recorded as soon as you click on the green microphone button. The button will turn red, indicating a sound is being recorded, and time remaining will show as the button "emptying it-self from its red liquid ". The data will then be processed, the digit coloring with change and the button will turn green again for you to input your next sound.*

Remember, you decide what sounds to associate to yellow and to grey. It is arbitrary and up to you. For example, the word "banana" could mean yellow, and "chocolate" mean grey. It will also work with non-words sounds, clapping your hands could mean yellow and snapping your finger could mean grey. You decide. A safe place to start is to say "yellow" for yellow and the "grey" for grey, boring but easy to remember.

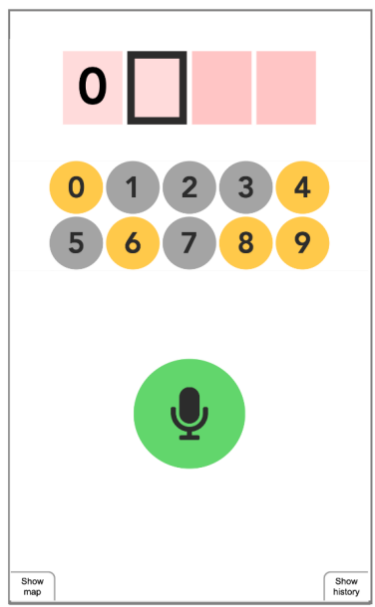
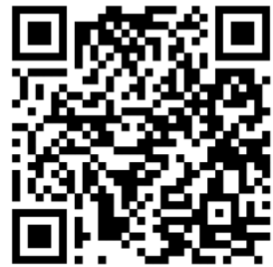

*Interaction 5: The self-calibrating version of IFTT-PIN using sounds as input. The user makes arbitrary sounds (e.g., spoken words, handclapping, etc.) and decides which sound means yellow and grey in their mind. Demo available at https://openvault.jgrizou.com/#/ui/demo_audio.json*

As for the sketch's version, the two buttons at the bottom of the screen allow you to see the projection of each sound. You can replay each sound by clicking on the play button. A video walkthrough is available via Video 8 below.

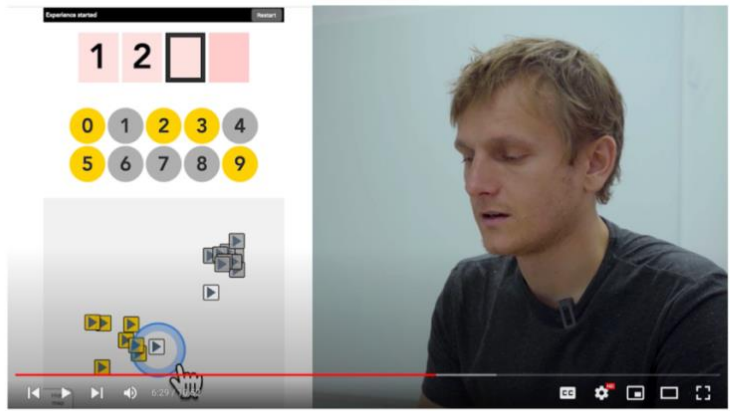
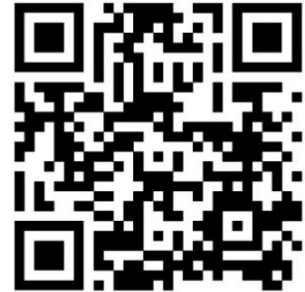

*Video 8: A step-by-step explanatory video of the self-calibrating version of IFTT-PIN using sounds as input. Video available at https://youtu.be/tiyQEdlu9RQ*

To visualize the process, you can use the interface with the dashboard as before, see Explanation 5. The side panel will show you the 2D representation of your label according to each hypothesis. It looks the same as in section 3.

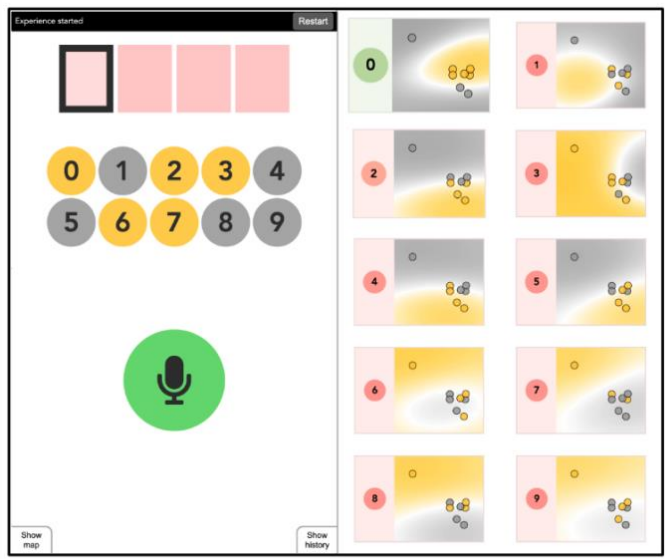
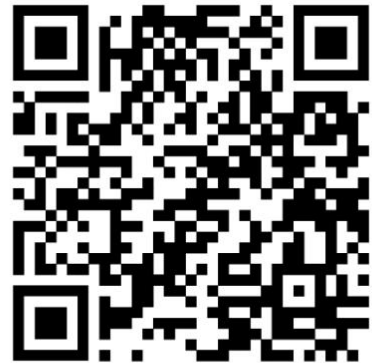

*Explanation 5: The self-calibrating version of IFTT-PIN using sounds as input and with a side explanatory panel. Demo available at https://openvault.jgrizou.com/#/ui/tuto_audio.json*

I included the sketch and sounds demonstrations to illustrate the potential of the SELF-CAL method to work with various real-world modalities. Once again, I invite you to reflect on how it feels to be able to decide how to communicate to a machine on the fly at interaction time. Whether it can be of practical use is another story of course but the increased agency it provides towards a digital tool is, to me, an interesting experience.

---

In the next section, we review domains where self-calibration problems have been encountered and discuss how other researchers approached this problem. We then list some open research questions and discuss potential applications along with ethical considerations.

# 6. Discussions

This section will be written later and discuss parallels I see with works in various fields as well as open challenges. You should not need this to be able to explore the content above and form your own opinion.

## 6.1. Related Work

## 6.2. Open Questions

## 6.3. Applications

# 7. Opening

I will update this section later discussing: "How would a world where interfaces are self-calibrating by default look like?"

# 8. Resources

All demos at [https://openvault.jgrizou.com/](https://openvault.jgrizou.com/)

Published papers:

- [https://arxiv.org/pdf/2205.09534.pdf](https://arxiv.org/pdf/2205.09534.pdf)
- [https://arxiv.org/pdf/2204.02341.pdf](https://arxiv.org/pdf/2204.02341.pdf)
- [https://arxiv.org/pdf/1906.02485.pdf](https://arxiv.org/pdf/1906.02485.pdf)
- [https://hal.archives-ouvertes.fr/hal-00984068/PDF/grizou2014calibration.pdf](https://hal.archives-ouvertes.fr/hal-00984068/PDF/grizou2014calibration.pdf)
- [https://hal.archives-ouvertes.fr/hal-01007689/PDF/grizou2014interactive.pdf](https://hal.archives-ouvertes.fr/hal-01007689/PDF/grizou2014interactive.pdf)
- [https://journals.plos.org/plosone/article?id=10.1371/journal.pone.0131491](https://journals.plos.org/plosone/article?id=10.1371/journal.pone.0131491)
- [https://hal.archives-ouvertes.fr/hal-00850703/document](https://hal.archives-ouvertes.fr/hal-00850703/document)
- [https://github.com/jgrizou/thesis_manuscript/releases/download/final/thesis.pdf](https://github.com/jgrizou/thesis_manuscript/releases/download/final/thesis.pdf)

# 9. Tables

## Table of Interactions

## Table of Explanations

## Table of Videos

# Table of Figures